\documentstyle[prl,aps,amsfonts,preprint,epsfig,floats]{revtex}

\tighten

\newcommand{\dsl}[1]{#1 \hspace{-0.15cm}\slash }
\newcommand{\imag}[0]{i}
\newcommand{\nrmv}[1]{| \vec{#1} |}

\newcommand{\Pslash}{P\!\!\!\!/\,}

\newcommand{\bp}{\begin{pmatrix}}
\newcommand{\ep}{\end{pmatrix}}
\newcommand{\Sc}{\scriptstyle}
\newcommand{\fourint}[1]{\int\!\frac{d^4 #1}{(2\pi)^4}}
\newcommand{\Slash}[1]{#1 \hspace{-.5em}/\hspace{.11em}}

\begin{document}
\date{\today}
\title{
\hspace*{\fill}{\small\sf UNITU--THEP--18/2000}\\
\hspace*{\fill}{\small\sf http://xxx.lanl.gov/abs/hep-ph/0012282}\\ ~\\ ~\\
Production Processes as a Tool to Study\\
Parameterizations of Quark Confinement}

\author{S.~Ahlig,
R.~Alkofer\footnote{E-Mail: reinhard.alkofer@uni-tuebingen.de},
C.~Fischer, M.~Oettel\footnote{Address after Feb. $1^{\rm st}$, 2001:
CSSM, University of Adelaide, SA 5005, Australia}, 
H.~Reinhardt, and H.~Weigel\footnote{Heisenberg--Fellow}}
\address{Institute for Theoretical Physics, University of T\"ubingen \\
          Auf der Morgenstelle 14, D-72076 T\"ubingen, Germany}

\maketitle

\begin{abstract}
We introduce diquarks as separable correlations in the two--quark 
Green's function to facilitate the description of baryons as relativistic
three--quark bound states. These states then emerge as solutions of 
Bethe--Salpeter equations for quarks and diquarks that interact via
quark exchange. When solving these equations we consider various 
dressing functions for the free quark and diquark propagators that
prohibit the existence of corresponding asymptotic states and thus 
effectively parameterize confinement. We study the implications of 
qualitatively different dressing functions on the model predictions
for the masses of the octet baryons as well as the electromagnetic 
and strong form factors of the nucleon. For different dressing 
functions we in particular compare the predictions for kaon 
photoproduction, $\gamma p\to K\Lambda$, and associated strangeness 
production, $pp\to pK\Lambda$,  with experimental data. This leads
to conclusions on the permissibility of different dressing functions.
\\
\end{abstract}
~

\newpage 

\section{Introduction}

The complexity of Quantum Chromodynamics (QCD) inhibits the computation of
hadronic properties and reactions from first principles. As a  consequence
models that potentially imitate the essentials of the  QCD dynamics have been
developed in the past to describe hadrons.  A relativistic description of
baryons as three--quark bound states is  provided by the solutions of the
Bethe--Salpeter equations\footnote{For further details on the application of
the Bethe--Salpeter formalism to QCD we refer to
reviews~\cite{Roberts:2000aa,Alkofer:2000wg}   and references therein.} for
quarks  and diquarks which interact via quark 
exchange~\cite{Alkofer:1995mv,Reinhardt:1990rw,Cahill:1989dx}. Once the full
three--quark  problem has been reduced to an effective two--body problem, the
only model  ingredients are the (di)quark propagators along with the
quark--diquark  vertices. It is hoped for that further progress in the study 
of the QCD quark propagator and two--quark correlations will eventually 
justify the reduction to quarks and diquarks in this approach to describe 
baryons. 

Actual calculations utilize either simplifying assumptions or 
phenomenological parameterizations of the respective propagators and 
interaction vertices of quarks and diquarks. By choosing the simplest
{\em ans\"atze}, {\em i.e.} free spin--1/2 and spin--0/spin--1 
propagators for quarks and diquarks, respectively, various spacelike 
nucleon form factors have been successfully reproduced~\cite{Oettel:2000jj}. 
However, the na{\"\i}ve use of perturbative (di)quark propagators leads 
to asymptotic states in the spectrum that carry the respective quantum 
numbers. Hence baryons would decay into quarks unless kinematically
bound. This decay process would contradict the confinement phenomenon.
In this paper we will therefore investigate the possibility of 
incorporating confinement into the diquark model by suitable 
modifications of the quark and diquark propagators. Essentially these
propagators are modified by multiplicative dressing functions to
completely remove the poles that occurred in the perturbative 
propagators at the (di)quark masses. This enables us to 
calculate the spectrum not only of octet but also decuplet 
baryons~\cite{Oettel:1998bk}. Together with an {\it ansatz} for the  
quark--diquark bound state wave--function of the nucleon (Faddeev 
amplitude) such pole--free propagators have already been used to 
calculate nucleon form factors in the spacelike 
regime~\cite{Bloch:2000rm,Bloch:1999ke}. Unfortunately in this 
context the computation of the electro--weak form factors is not 
as simple as merely modifying the propagators. Since gauge 
invariance relates off--shell propagators and vertices it is obvious 
that any change in the propagators requires modifications of the 
effective interaction with the electro--weak gauge 
bosons~\cite{Blankleider:2000xp}. When incorporating gauge invariance 
in the model with free propagators the nucleon isovector magnetic 
moments come out too small by about 30\% unless
model parameters are used that do not properly reproduce the baryon
spectrum~\cite{Oettel:2000jj}. However, these unacceptable parameters
result from requiring the $\Delta$--isobar to be kinematically
bound against its decay into free quarks. It is hoped for that 
when modeling confinement the results on the magnetic moments will
also improve due to the modifications of the photon vertices which are
mandatory when employing dressed (di)quark propagators. 
A very serious disadvantage of the lack of confinement is that hadronic 
reactions with {\em timelike} momenta of the order of $1{\rm GeV}$ 
transferred to the nucleon, {\it e.g.} meson production processes, cannot 
be described properly. Again, the free--particle poles of quark and 
diquark cause unphysical thresholds in these processes that are triggered
by the poles in the propagators. An appropriate modification
of these propagators would not only remove the unphysical
thresholds but also serve as an effective description of the strong
interaction. Certainly, a relativistic description of such processes 
would be very desirable. At present, the covariant diquark model 
appears to be the only relativistic one that is both, feasible and 
applicable at this energy scale. 

As already mentioned we wish to eliminate the singularities associated with
real timelike momenta in the (di)quark propagators that would lead  to
imaginary parts in those $S$--matrix elements that are calculated from 
diagrams containing internal quark loops. So, either these singularities are
absent or their contributions cancel in some manner~\cite{Alkofer:2000wg}. 
The qualitative behavior described can be encoded in the following
models (which are certainly not the only possibilities) for the quark 
propagator in Euclidean space,
\begin{eqnarray}
 S^{(k)} (p) &=& \frac{\imag \dsl{p}-m_q }{p^2+m_q^2}  \;
   f_{k}\left(\frac{p^2}{m^2_q}\right) \; ,
  \qquad k = 0,\ldots,3 \; ,
\label{sk}
\end{eqnarray}
with
\begin{eqnarray}
  f_0 (x) &=& 1 \hspace{2cm} \text{(bare propagator)}\;
    \label{f0},\\
  f_1 (x) &=& \frac{1}{2}\left\{
  \frac{x+1}{x+1-i/d} + \frac{x+1}{x+1+i/d} \right\}
    \label{f1} \; ,\\
  f_2 (x) &=&
   1- \exp\left[ -d\left(1+x \right) \right]
    \label{f2}\; , \\
  f_3 (x,x^\ast) &=& \tanh\left[ d\left(1+x\right)
         \left(1+x^\ast\right) \right] \; .
 \label{f3}
\end{eqnarray}
The propagator~(\ref{f1}) possesses complex conjugate 
poles~\cite{Stingl:1996nk} such that corresponding virtual 
excitations cancel each other in physical amplitudes. Here $m$ represents 
a parameter that would be interpreted as the quark mass if and only if 
the poles were on the real axis. In another 
scenario~(\ref{f2}), the dressing functions are chosen such that the
propagators are entire  functions and non--trivial in the whole complex
plane~\cite{Efimov:1993zg,Bloch:1999ke}. If they are required to be analytic,
they must possess an essential singularity, at least for infinite arguments.
Third, it might be helpful to approximate propagators by non--analytic
functions~(\ref{f3}) and constrain them such that they 
asymptotically behave like $1/|p|^2$ for both, large spacelike and 
timelike momenta. Since we enforce the propagators to be free of poles, 
they must be non--analytic functions depending on both the particle 
momentum $p$ and its complex conjugate~$p^\ast$.  Consequently the 
quark--photon and quark--meson vertices are non--analytic and, 
even worse, translation invariance is lost in the solutions to 
the nucleon Bethe--Salpeter equation. These issues will be detailed 
in section~\ref{dqmodel}. The trivial dressing function $f_0$ 
corresponds to the bare propagator. We will consider this case only 
for comparison.

\begin{figure}[t]
  \begin{center}
  \epsfig{file=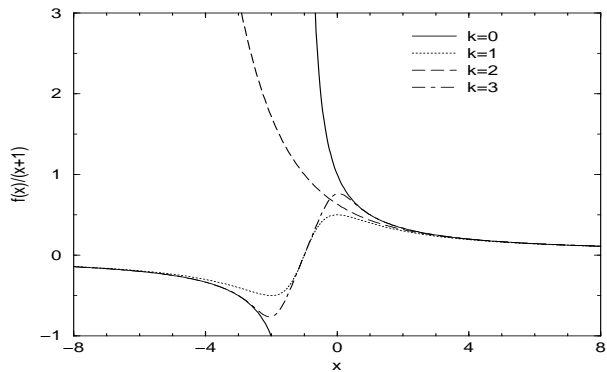,width=8cm,height=5cm}
  \caption{\sf The propagator functions,
           $\tilde{f}_k(x)=f_k (x)/(x+1)$ for real $x$ and
           for $k=0,\ldots,3$, {\it cf.} eqs~(\ref{f0})--(\ref{f3}).
           The thick solid line corresponds to the free propagator.
           Here we have set $d=1$.}
  \label{f123_re}
  \end{center}
\end{figure}
In figure~\ref{f123_re} we show $\tilde{f}_k(x)=f_k(x)/(x+1)$ for
$k=0,\ldots,3$ for real $x$. Note that these dressing functions
are real in that case. We observe that $\tilde{f}_1(x)$ and
$\tilde{f}_3(x)$ change sign (as in the case of a tree--level propagator)
while the function $\tilde{f}_2(x)$ increases drastically.
For asymptotically large spacelike momenta the three model propagators
$S^{(k)},\, (k=1,2,3)$ match up with the bare propagator $S^{(0)}$.
Our present investigation focuses on the phenomenological implications 
of the so--modified propagators.

This paper is organized as follows: In Section \ref{dqmodel} the covariant 
diquark model for baryons is presented. The corresponding  
Bethe--Salpeter equation that describes baryons as bound states of quarks and
diquarks is derived in appendix~\ref{3qreduce}. 
The formalism of refs.~\cite{Oettel:2000gc,Oettel:2000jj}
for calculating form factors is described for later determination
of model parameters.
Using the above given scenarios for implementing the
confinement phenomena at the level of propagators this will set the stage for
the main topic of our paper: The sensitivity of  the predicted observables on
the various effective parameterizations of  confinement. These
parameterizations concern the structure of the  (di)quark propagators for
complex momenta. In section~\ref{relevance}  we will discuss the regime of
complex momenta that is relevant for studying the baryon spectrum as well as
several production processes. In section~\ref{CalcObserv} we will describe the
formalism necessary  to compute various production processes in the
diquark--quark model. These  comprise especially the cross sections for kaon
photoproduction  $p\gamma \rightarrow K\Lambda$ and the associated strangeness
production  in $pp \rightarrow pK\Lambda$. We will proceed by presenting our
numerical  results in section~\ref{num_res}, including the determination of the
model parameters. In particular we will compare the  predictions that originate
from the different dressing functions for the  propagators. Finally, we will
conclude by formulating criteria for phenomenologically permissible
parameterizations of the propagators. Some derivations and technical details
are relegated to four appendices.

\section{The Covariant Diquark Model for Baryons}
\label{dqmodel}

\subsection{The diquark--quark Bethe--Salpeter equation}
\label{BSE-sec}

\begin{figure}[b]
 \begin{center}
   \epsfig{file=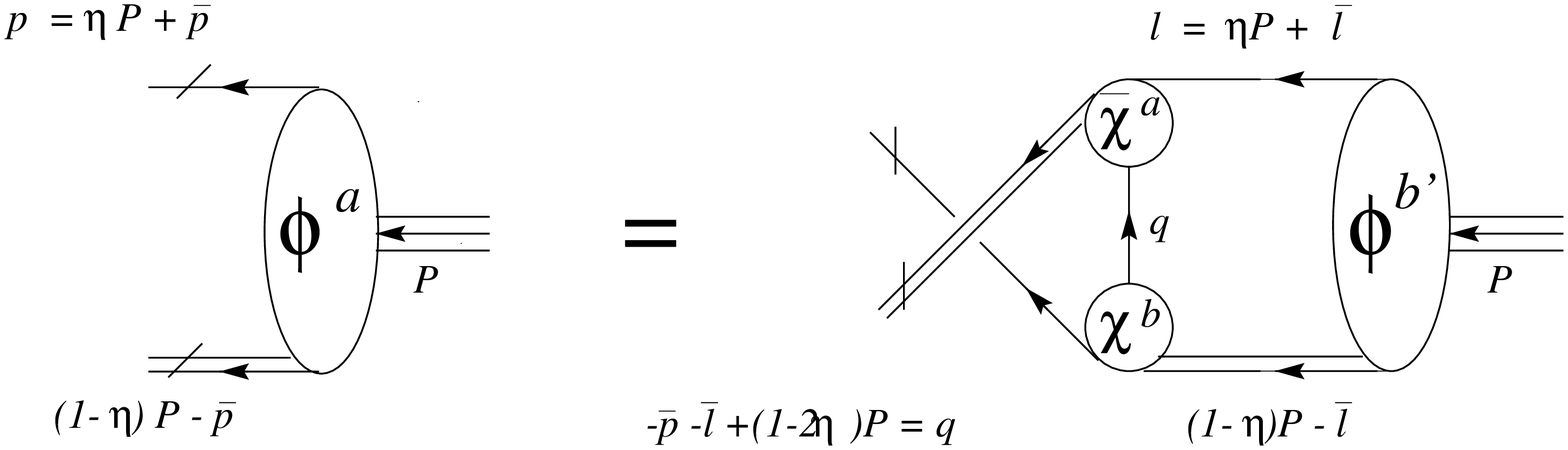,width=8cm,height=2.2cm}
 \end{center}
 \caption{\sf The coupled set of Bethe-Salpeter equations
          for the effective vertex functions $\phi^a$.}
 \label{bse_fig}
\end{figure}

We start from the Faddeev formalism for three quarks and impose two 
essential assumptions to arrive at a Bethe--Salpeter equation
that describes baryons as bound states of quarks and diquarks 
interacting via quark exchange. These assumptions are (i) all 
three--particle irreducible graphs can be safely omitted and (ii)
the two--quark correlations can be approximated by separable
correlations, the so--called diquarks. The actual derivation of 
the Bethe--Salpeter equation for the effective baryon--diquark--quark
vertex functions $\phi^a$ is presented in appendix~\ref{3qreduce}. 
Using the definitions for total and relative momentum given as in 
figure~\ref{bse_fig} this integral equation reads
\begin{eqnarray}
\phi^a_{i,\alpha}(\bar{p}_i,P)& =& \sum_{bb^\prime} 
\int \frac{d^4\bar{l}}{(2\pi)^4} \; 
K^{ab}_{ij,\beta\alpha}(\bar{p}_i,\bar{l},P)\;
G_{j,bb^\prime}^{\beta\beta^\prime}(\bar{l},P)\;
\phi^{b^\prime}_{j,\beta^\prime}(\bar{l},P) \;
\,+\, \left(j\longleftrightarrow k\right)\, .
   \label{BS2}
\end{eqnarray}
Here $G_j$ describes the disconnected quark--diquark 
propagator 
\begin{equation}
G_{j,bb^\prime}^{\beta\beta^\prime}(\bar{l},P)=
S_j^{\beta\beta^\prime}(\eta P+\bar{l})\; 
D_{bb^\prime}((1-\eta)P-\bar{l}) \; = \;
S_j^{\beta\beta^\prime}(l)D_{bb^\prime}(P-l)
\label{Gqdq}
\end{equation}
Furthermore, the quark--diquark interaction kernel $K$ contains besides 
the propagator of the exchanged quark also the diquark amplitudes $\chi_i^a$
defined via the separability assumption 
\begin{equation}
t_i(k_j,k_k;p_j,p_k)=\sum_{a,a^\prime} \chi_i^a(k_j,k_k)\;
D_{a,a^\prime}(k_j+k_k)\; \bar\chi_i^{a^\prime}(p_j,p_k)\; 
\end{equation}
of the quark--quark 
$t$--matrix (cf. App. \ref{3qreduce}). The kernel explicitly reads
\begin{eqnarray}
K^{ab}_{ij,\beta\alpha}(\bar{p},\bar{l},P)&=&
\bar{\chi}^a_{i,\beta\gamma}(\bar{l}+\eta P,q) \;
S_k^{\gamma\gamma^\prime}(q)\; 
\chi^b_{j,\gamma^\prime\alpha}(q,\bar{p}+\eta P)  
\nonumber \\*
&=&\bar{\chi}^a_{i,\beta\gamma}(l,q)\;S_k^{\gamma\gamma^\prime}(q)\;
\chi^b_{j,\gamma^\prime\alpha}(q,p)\; ,
\label{KBS} \\*
{\rm with}&& \nonumber \\*
q&=&(1-2\eta)P-\bar{p}-\bar{l} \; =\; P-p-l \, ,
\nonumber
\end{eqnarray}
since $P=p_i+p_j+p_k$ and $\bar{l}=l-\eta P$. The above relations 
also indicate the independence of the momentum partition parameter $\eta$ 
since the Jacobian of the transformation $\bar{l}\to l$ equals unity 
for fixed total momentum $P$. 

For the solution of the Bethe--Salpeter equation~(\ref{BS2}) we still
have to choose the appropriate quantum numbers associated with
baryons. This will be discussed in subsection~\ref{decomposition}
and we will find that the quark exchange (parameterized by the kernel 
$K^{ab}$) generates sufficient attraction to bind quarks and diquarks 
to baryons. For identical quarks antisymmetrization is required when
projecting onto baryon quantum numbers. Fortunately, this does not alter 
the algebraic form of the Bethe--Salpeter equation~(\ref{BS2}). Rather,
it simply implies that we may omit the single particle indices $i$ on 
the quark propagators $S_i$. Only when caring about the discrete quantum 
numbers we have to revert to these indices since they specify the 
{\em summation order} over color, flavor and Dirac indices in 
eq.~(\ref{BS2}). Furthermore the functional forms of the 
diquark propagators $D_{aa^\prime}$ and the vertices $\chi^a_i$ do 
not depend on the quark labels. These independencies are already indicated 
in eqs~(\ref{Gqdq}) and~(\ref{KBS}) as we have omitted the quark labels for 
the momenta. 

In a self--consistent approach one would calculate the $t$--matrix from its own
Bethe--Salpeter equation~(\ref{hatt_i}). However, this is beyond the scope of the 
present investigation. Instead we model the $t$--matrix by diquark correlators
which have an analytic structure such that no particle interpretation for the
diquark exists.  
We will restrict ourselves to the scalar and axialvector channels
as these comprise the minimal set to describe octet and decuplet baryons.
Furthermore these channels are generally assumed to be the most important 
ones, see refs.~\cite{Roberts:2000aa,Alkofer:2000wg} and references therein. 
The corresponding separable {\it ansatz} for the two--quark $t$--matrix reads
\begin{eqnarray} 
 t_{\alpha\beta,\gamma\delta}(k_1,k_2;p_1,p_2) &=&
 \chi^5_{\alpha\beta}(\bar{k},P_2) \;D(P_2)\;\bar
  \chi^5_{\gamma\delta}(\bar{p},P_2) \; +\; 
\chi_{\alpha\beta}^\mu(\bar{k},P_2) \;D^{\mu\nu}(P_2)\;
    \bar  \chi_{\gamma\delta}^\nu(\bar{p},P_2)  \; .
  \label{tsep}
\end{eqnarray}
Here we rewrite the diquark--quark vertices $\chi^{5[\mu]}$ as functions
of relative,$\bar{k}=\sigma k_1 -(1-\sigma)k_2$ , and total,
$P_2=k_1+k_2=p_1+p_2$ , momenta instead of the single quark momenta.
In actual calculations, we 
choose for simplicity the symmetric momentum partition, {\it i.e.} 
$\sigma=1/2$. Shifting the value of $\sigma$ is possible, however, 
this complicates
slightly the parameterization of diquark correlations, see the discussion 
below eq.\ (\ref{dqvertex_a}) and in ref.~\cite{Oettel:2000gc}.

The diquark propagators in the scalar and the axialvector channel
are modeled as
\begin{eqnarray}
D(P) &=& -\frac{1}{P^2+m_{sc}^2}\, f\left(\frac{P^2}{m^2_{sc}}\right) \; ,
 \label{Ds} \\
D^{\mu\nu}(P) &=& -\frac{1}
{P^2+m_{ax}^2} \left( \delta^{\mu\nu}+(1-\xi) \frac{P^\mu P^\nu}{m_{ax}^2}
\right)\,f\left(\frac{P^2}{m^2_{ax}}\right)  \; . \quad
 \label{Da}
\end{eqnarray}
The dressing function $f(P^2/m^2)$ is hereby chosen identical to the one for
the quark propagator, i.e.\ either one of the forms
(\ref{f1},\ref{f2},\ref{f3}). Note that the choice $f(P^2/m^2)=1$  and $\xi=0$
corresponds to the free propagators of spin--0 and spin--1  particles. As a
major purpose of the present paper we will study  various deviations from the
free propagators as an avenue to mimic confinement. In general, the dressing
functions $f$ are different in the scalar and axialvector channels as well as
they are distinct from the one for the quark propagator. For simplicity,
however, we will assume identical functions for all propagators. As we will
not consider any axialvector diquark loops it is sufficient for the present
purpose to use $\xi=1$, see ref.\ \cite{Oettel:1998bk} where it has been shown
that choosing $\xi=1$ leads to almost identical results for baryon amplitudes as
$\xi=0$.

If diquark poles existed in the $t$ matrix, the diquark--quark vertices  $\chi$
and $\chi^\mu$ would on-shell \mbox{$(P^2=-m^2_{sc[ax]})$}  correspond to
diquark Bethe--Salpeter vertex functions. These vertex functions have a finite 
extension in momentum space and fall off fast
enough to render all integrals finite.
Empirically  we
assume that the corresponding scale is linked to the (inverse)  proton radius.
The conjugate vertex functions $\bar{\chi}$ are obtained  by charge
conjugation,
\begin{eqnarray}
 \bar \chi^5 (p,P)& =& C\; \left( \chi^5 (-p,-P) \right)^T
      \; C^T \;, \label{dscon} \\
 \bar \chi^{\mu} (p,P)& =& -C\; \left( \chi^{\mu} (-p,-P) \right)^T
      \; C^T \;, \label{dacon}
\end{eqnarray}
where $T$ denotes the transpose. 

Let us now explicitly construct the vertex functions. They must be
antisymmetric under the interchange of the two quarks. This entails
\begin{equation} \chi^{5[\mu]}_{\alpha\beta}(\bar{p},P) = \left.
-\chi^{5[\mu]}_{\beta\alpha}(-\bar{p},P) \right|_{\sigma \leftrightarrow
(1-\sigma)} \; . \label{antisymm1} \end{equation} Any two quarks within a
baryon belong to the color antitriplet representation. Thus the diquark--quark
vertices are proportional to the antisymmetric tensor $\epsilon_{ABD}$. Here
$A$ and $B$ are the color indices of the quarks whereas $D$ labels the color of
the diquark.  Furthermore the scalar diquark is antisymmetric while the
axialvector  diquark is symmetric in flavor. We maintain only the dominant
components with regard to the structure in Dirac space.\footnote{The complete
Dirac structure for the scalar diquark containing four independent tensors  
can be obtained by analogy from the one for pseudoscalar mesons.
The complete Dirac structure for the axialvector diquark
consists of twelve independent terms, four longitudinal and eight transverse
ones.} These are the antisymmetric matrix $(\gamma^5 C)$ for the scalar diquark
and  the symmetric matrices $(\gamma^\mu C)$ for the axialvector diquark.
Considering, for the time being, only two flavors the vertices then 
read\footnote{In the compact notation the indices  $\alpha$ and $\beta$ of
$\chi_{\alpha\beta}$ not only contain  the Dirac labels but also those for
flavor and color.} \begin{eqnarray}
\chi^5_{\alpha\beta}(\bar{p},P)\big|_{\sigma=1/2}=
\chi^5_{\alpha\beta}(\bar{p})&=&g_s (\gamma^5 C)_{\alpha\beta}\;  V(\bar{p}^2)
\;\; \frac{(\tau_2)_{ab}} {\sqrt{2}}\,\frac{\epsilon_{ABD}}{\sqrt{2}} \;,
\label{dqvertex_s} \\ \chi_{\alpha\beta}^{\mu}(\bar{p},P)\big|_{\sigma=1/2}=
\chi_{\alpha\beta}^{\mu}(\bar{p})&=&g_a (\gamma^\mu C)_{\alpha\beta}\; 
V(\bar{p}^2) \;\; \frac{(\tau_2\tau_k)_{ab}}
{\sqrt{2}}\,\frac{\epsilon_{ABD}}{\sqrt{2}}\; . \label{dqvertex_a}
\end{eqnarray} Choosing the scalar function $V$ to depend only on the squared
relative  momentum $\bar{p}^2$, these vertices are indeed antisymmetric with
respect  to exchange of quark labels for the partition $\sigma=1/2$.  Otherwise
a parametrization of
$V$ would depend on both $\bar{p}^2$ and  $\bar{p}\cdot
P$ in order to comply with antisymmetrization~\cite{Oettel:2000gc,Maris:1997tm}.
However, complete independence for observable quantities on
$\sigma$ could only be obtained by solving the Bethe--Salpeter equation
(\ref{hatt_i}) for the two--quark $t$--matrix in which case the scalar functions
$V$ could depend on the quantity $(\bar p \cdot P)^2$ (for $\sigma=1/2$) 
which is symmetric under quark exchange.
In the actual calculations we will use a multipole
form type {\it ansatz} \begin{eqnarray} V(x)=V_n(x) & =&
\left(\frac{\lambda_n^2}{\lambda_n^2 + x} \right)^n \; .\label{npole}
\end{eqnarray}     

The overall strength of the diquark correlations given in
eqs~(\ref{dqvertex_s},\ref{dqvertex_a}) is governed by the 
``diquark-quark coupling constants'' $g_s$ and $g_a$. 
They could be determined by either imposing the canonical Bethe--Salpeter
norm condition on $\chi^{5[\mu]}$ or by the solution to the differential
Ward identity for the diquark--photon vertex which is sensitive to the 
substructure of the diquarks \cite{Oettel:2000jj}. For simplicity, we 
will fix $g_s$ from fitting the nucleon mass. When 
including axialvector diquarks we will assume the ratio $g_a/g_s=0.2$
as suggested by the results of ref.~\cite{Oettel:2000jj}. In this
manner the baryon Bethe--Salpeter equation~(\ref{BS2}) becomes
an eigenvalue problem for the coupling constants $g_s$ and $g_a$.

Note that by parameterizing the quark--quark $t$--matrix we do not
make any reference to the nature of the relevant quark--quark 
interaction. For example, to quantitatively include pionic effects
we would have to solve Dyson--Schwinger equations for 
the quark propagator and the Bethe--Salpeter equation with explicit
pion degrees of freedom. Studies within the Nambu--Jona--Lasinio 
model using diquark--quark correlations either in a soliton 
background~\cite{Zuckert:1997nu} or with explicit pion interaction
between the quarks~\cite{Ishii:1998tw} lead to a substantial gain 
in the binding energy. Since we determine the coupling constant 
$g_s$ from the nucleon mass, those studies suggest that the inclusion
of pion degrees of freedom would merely lead to a shift of this
constant.

Equipped with the separable form of the two--quark correlations,
see eq~(\ref{tsep}), and the functional form of the scalar and axialvector 
diquark correlations in eqs~(\ref{dqvertex_s},\ref{dqvertex_a}), we will
set up the effective Bethe-Salpeter equation for the nucleon.

Upon attaching quark and diquark legs to the vertex functions $\phi$
one obtains the Bethe--Salpeter wave functions $\psi$.
Equation (\ref{BS2}) can then be rewritten as a system of equations for 
wave-- and vertex functions as defined in Appendix \ref{decomposition}.
Using the notations (\ref{vertnucdef}--\ref{wavenucdef}) we obtain
\begin{equation}
  \fourint{k} G^{-1}(p,k,P)
  \pmatrix{\Psi^5(k,P) \cr \Psi^{\mu'}(k,P)}=0 \;.
  \label{bse_nuc}
\end{equation}
Here $G^{-1}(p,k,P)$ is the inverse of the quark--diquark four--point
function which results from the quark exchange\footnote{For convenience
we have omitted the discrete labels.}. It is the sum of the disconnected 
part and the interaction kernel which contains the quark exchange,
\begin{eqnarray}
 G^{-1} (p,k,P) &=&
    (2\pi)^4 \;\delta^4(p-k)\; S^{-1}(p_q)\;
       \circ  D^{-1}(p_d) \\
 & &\mbox{\hskip 1.0cm}-\frac{1}{2}
\pmatrix{-\chi^5{\Sc (p_2^2) } \; S^T{\Sc (q) }\; 
           \bar\chi^5{\Sc (p_1^2) } &
          \sqrt{3}\; \chi^{\mu'}{\Sc (p_2^2) }\; S^T{\Sc (q) }\;
          \bar\chi^5 {\Sc (p_1^2) } \cr
    \sqrt{3}\;\chi^5{\Sc (p_2^2) }\; S^T{\Sc (q) }\;
      \bar\chi^{\mu}{\Sc (p_1^2) } &  
     \chi^{\mu'}{\Sc (p_2^2) }\; S^T{\Sc (q) }\;
     \bar\chi^{\mu}{\Sc (p_1^2)}
} . \nonumber
\end{eqnarray}
The flavor and color factors have been worked out and therefore 
$\chi^5(p^2)$ and $\chi^{\mu}(p^2)$ from now on only represent the Dirac structures 
of the diquark--quark vertices (multiplied by the invariant function
$V_n(p^2)$, {\it cf.} eq~(\ref{npole})).
The freedom to partition the total momentum between quark and diquark
introduces the parameter $\eta \in [0,1]$ with $p_q=\eta P+p$ and
$p_d=(1-\eta)P - p$. The momentum of the exchanged quark is then given by
$q=-p-k+(1-2\eta)P$. The relative momenta of the quarks in the diquark
vertices  $\chi$ and  $\bar\chi$ are $p_2=p+k/2-(1-3\eta)P/2$ and
$p_1=p/2+k-(1-3\eta)P/2$, respectively. Invariance under (four dimensional) 
translations implies that for every solution $\Psi(p,P;\eta_1)$ of the 
Bethe--Salpeter equation a family of solutions exists that have the form 
$\Psi(p+(\eta_2-\eta_1)P,P;\eta_2)$. Considering the Bethe--Salpeter 
equation as a linear eigenvalue problem for $\Psi$ (or $\Phi$) in the 
coupling constant $g_s$, translation invariance requires
the coupling constant eigenvalue to be independent of $\eta$ once a 
bound--state mass $-P^2=M^2$ is fixed. 
This independence is exactly what one observes in the numerical solutions
of the BSE, provided the analytic form of the dressing functions, 
eq~(\ref{f0})--(\ref{f2}), is used.
However, the $\eta$--independence is 
lost when substituting non--analytic propagators such as those 
parameterized by the function $f_3$, eq~(\ref{f3}). Essentially the 
reason is that Cauchy's theorem does not apply to non--analytic functions. 
The difference in the eigenvalues of the Bethe--Salpeter equation
under the variation of $\eta$ can be shown to equal a contour integral 
in the complex $p$--plane. This integral vanishes only if the 
integrand is an analytic function. However, when choosing $d>5$ in 
eq~(\ref{f3}), the propagator resembles the free propagator in a 
large domain thereby mitigating the $\eta$--dependence.

The structure of the equations for the octet baryons is similar to
that of the nucleon~(\ref{bse_nuc}). However, the number of Dirac 
structures $\Phi^5$ and $\Phi^\mu$ increases due to the possible different 
quark--diquark flavor configurations. These equations are given in full 
detail in refs.~\cite{Oettel:1998bk,Oettel:diss}. 
Allowing for flavor symmetry breaking, that is induced by a difference 
between the masses of strange quark and up/down quark, discriminates
vertex functions $\Phi^5$ and $\Phi^\mu$ with different diquark 
configurations~\cite{Oettel:1998bk}. As the $\Lambda$ hyperon presently 
is of special interest, we list its three different correlations,
\begin{equation}
 \Phi_\Lambda \sim  \Phi^5_{F_1}\,, \quad
 \Phi^5_{F_2} \quad {\rm and} \quad \Phi^\mu_L \; .
\end{equation}
Here, $F_1=\{d[us]-u[ds]\}/\sqrt{2}$, $F_2=s[ud]$ and 
$L=[d\{us\}-u\{ds\}]/\sqrt{2}$ refer to different quark--diquark flavor 
states. Antisymmetrized scalar diquarks are denoted
by square brackets $[\dots]$ and symmetrized axialvector diquarks 
by curly brackets $\{\dots\}$. Note that broken $SU(3)$--flavor symmetry 
induces a component of the total antisymmetric flavor singlet 
$\frac{1}{\sqrt{3}}\left[[su]d+[ud]s+[ds]u\right]$ 
into wave and vertex functions. 
In non--relativistic quark models with $SU(6)$ symmetry such a component is 
forbidden by the Pauli principle, however, having non--vanishing lower 
components in the baryon bi--spinors does actually lead to such flavor 
singlet components. In actual calculations they turn out to be small
~\cite{Oettel:1998bk}.

\subsection{Electromagnetic form factors}

\begin{figure}[t]
\begin{center}
 \epsfig{file=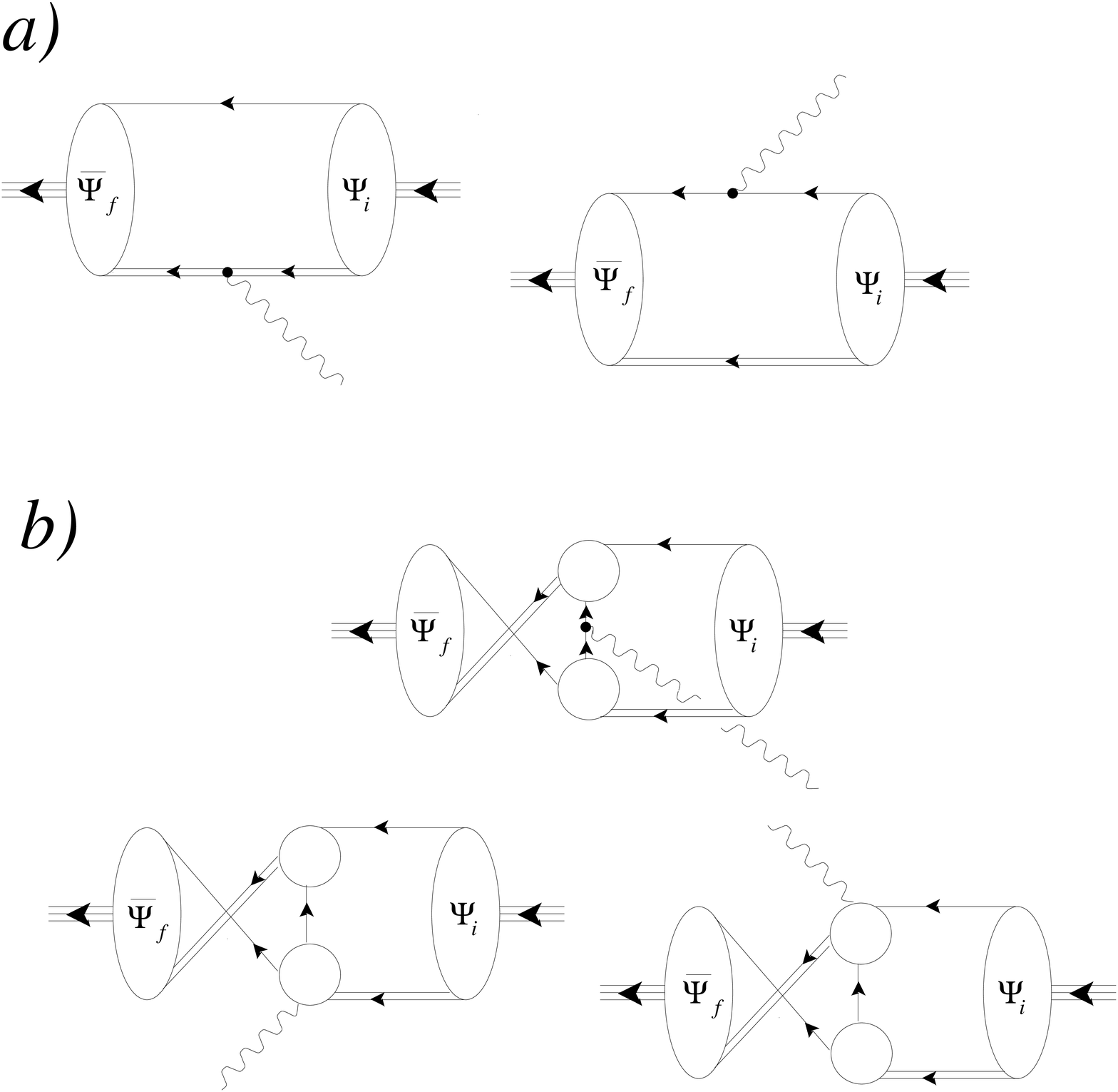,width=10cm,height=12cm}
\vskip0.5cm
\caption{\sf Contributions from the impulse approximation 
$(a)$ and from the Bethe--Salpeter kernel $(b)$ to the electromagnetic 
form factors.}
\label{emff}
\end{center}
\end{figure}

To further constrain the model parameters, we calculate 
the electromagnetic form factors of the nucleon. 
In this section we provide the formalism and the corresponding
results will be given in subsection \ref{para_fix}.
These form factors 
parameterize the nucleon matrix element of the current operator
that describes the coupling of the photon to quark and diquark within 
the nucleon. Gauge invariance and the proper normalization of the 
nucleon charges are ensured when the current operator comprises all 
possible couplings to the inverse quark--diquark four--point function 
$G^{-1}$ of eq~(\ref{bse_nuc})~\cite{Oettel:2000gc,Oettel:2000jj,Oettel:diss}. 
The current operator is sandwiched between the Bethe--Salpeter
wave--functions $\bar \Psi$ and $\Psi$ of the final and initial state,
respectively, according to Mandelstam's recipe~\cite{Man55}. In total, we have 
contributions from the {\em impulse approximation} which are described by 
the upper diagrams in figure~\ref{emff} and contributions from the 
{\em Bethe--Salpeter kernel} which are given by diagrams of the 
type given in the lower part of figure~\ref{emff}.

To calculate the form factor diagrams, we need properly normalized
Bethe--Salpeter wave--functions. This normalization is obtained from
\begin{eqnarray}
  M_N \Lambda^+ \; \stackrel{!}{=}
  -\int \frac{d^4\,p}{(2\pi)^4}
  \int \frac{d^4\,k}{(2\pi)^4} 
   \bar \Psi(k,P_n) \left[ P^\mu \frac{\partial}{\partial P^\mu}
    G^{-1} (k,p,P) \right]_{P=P_n} \hskip -.5cm  \Psi(p,P_n) \; ,
  \label{normnuc}
\end{eqnarray}
where $M_N$ is the nucleonic bound state mass. The conjugated 
wave--function $\bar \Psi$ is in analogy with eqs (\ref{dscon}) and
(\ref{dacon}) given by
\begin{equation}
\bar{\Psi}(k,P_n)=\eta_d\, C\,\Psi(-k,-P_n)^T\, C^T
\label{psi_con}
\end{equation}
with $\eta_d=1$ and $\eta_d=-1$ when the involved diquark is
respectively of scalar or axialvector type.

Furthermore we need expressions for the photon vertices that appear
in the diagrams. In ref.~\cite{Oettel:2000gc} the seagull vertices describing 
the photon coupling to the diquark--quark vertices $\chi$ have been 
derived for the scalar diquark. The coupling of the axialvector diquark
to the photon has been studied in ref.~\cite{Oettel:2000jj}.
The photon vertices with quark and diquark must fulfill the
differential Ward identities for zero momentum transfer to the photon
\begin{eqnarray}
 \label{diffq}
  \Gamma^\mu_q (p_q,p_q) &=& \frac{\partial}{\partial p_q^\mu}\;
    S^{-1}(p_q)\;, \\*
  \tilde \Gamma^\mu_d (p_d,p_d) &=&
 \label{diffd} 
    \pmatrix{\Gamma^\mu_s & 0 \cr 0 & \Gamma^\mu_a}
    = \frac{\partial}{\partial p_d^\mu}\;
    \tilde D^{-1}(p_d) \; .
\end{eqnarray}  
Here $\tilde \Gamma^\mu_d$ comprises both photon vertices with scalar 
and axialvector diquarks. For convenience the discrete labels have
been omitted. The Ball--Chiu construction of the longitudinal part of 
the vertices ensures that they obey both the differential Ward as 
well as the Ward--Takahashi identity. The latter identities constrain 
the vertices for finite $Q$. For the dressed quark propagators of 
eq~(\ref{sk}) the corresponding vertices finally read
\begin{eqnarray}
 \Gamma^\mu_{q,i}(k_q,p_q) &=&
    -\frac{i}{2}\,\gamma^\mu\left[1/f_i(k_q^2/m_q)
   +1/f_i(p_q^2/m_q) \right] \nonumber \\
  && -\frac{i}{2}\,(k_q+p_q)^\mu\,\frac{1/f_i(k_q^2/m_q) - 
   1/f_i(p_q^2/m_q) }{k_q^2-p_q^2}
   \left[ (\Slash{k}_q+\Slash{p}_q)-2im_q \right] \; .
\label{wtf012}
\end{eqnarray}
This construction is valid for the analytic dressing functions
$f_0,f_1$ and $f_2$. 

In the case of the non--analytic dressing function $f_3$, we must specify
the derivatives in eqs~(\ref{normnuc}--\ref{diffd}) with respect to the 
total bound state momentum $P$ and the quark and diquark momenta
$p_q$ and $p_d$, respectively. We calculate form factors in the 
Breit frame, {\it i.e.} the temporal component of the 
momentum transfer is zero. Consequently, the relative momenta 
between the initial quark, $k$ and the final diquark, $p$ are real. 
They must be integrated over in the norm integral (\ref{normnuc}) and 
in the calculation of the diagrams of figure~\ref{emff}. Let us consider 
the quark momenta which are defined as before: $k_q=\eta P_f+k$ 
and $p_q=\eta P_i+p$. We define the derivatives in 
equations~(\ref{normnuc}),~(\ref{diffq}) and (\ref{diffd}) as follows,
\begin{equation}
 \frac{\partial}{\partial P_{i[f]}}=\eta\frac{\partial}
{\partial p[k]} \; , \quad 
 \frac{\partial}{\partial p_q [k_q]}
=\frac{\partial}{\partial p[k]} \; .
\end{equation}
Of course, these are trivial identities when applied onto analytical
functions. Derivatives with respect to the diquark momenta are
defined accordingly. The nucleon charges obtained as the form factors
at zero momentum transfer are then properly normalized. The 
corresponding proof utilizes the methods outlined in
ref.~\cite{Oettel:2000gc}.

To comply with the Ward--Takahashi identity, the quark--photon
vertex has to be modified 
\begin{eqnarray}
 \Gamma^\mu_{q,3} &=&  -\frac{i}{2}\,\gamma^\mu
  \left(1/f_3(k_q^2/m_q^2,k_q^{\ast2}/m_q^2)
   + 1/f_3(p_q^2/m_q^2,p_q^{\ast2}/m_q^2) \right) \nonumber \\
  && -\frac{i}{2}\,(k_q+p_q)^\mu\,\frac{1/f_3(k_q^2/m_q^2,p_q^{\ast2}/m_q^2) -
   1/f_3(p_q^2/m_q^2,p_q^{\ast2}/m_q^2) }{k_q^2-p_q^2} 
  \left[ (\Slash{k}_q+\Slash{p}_q)-2im_q \right] \nonumber \\
  &&-\frac{i}{2}\,(k^\ast_q+p^\ast_q)^\mu\,
     \frac{ 1/f_3(k_q^2/m_q^2,k_q^{\ast2}/m_q^2) -
   1/f_3(k_q^2/m_q^2,p_q^{\ast2}/m_q^2)}{k_q\cdot k^\ast_q-p_q\cdot p^\ast_q} 
   \left[ (\Slash{k}_q+\Slash{p}_q)-2im_q \right] \; . 
\label{wtf3}
\end{eqnarray}
This vertex now depends on the four variables $k_q,k^\ast_q,p_q$ and
$p^\ast_q$ and is also non--analytic as is the corresponding quark
propagator $S^{(3)}$. The photon Ball--Chiu vertices with scalar and 
axialvector diquarks have to be modified using an analogous 
description\cite{Oettel:diss}. The coupling of the photon 
to the anomalous magnetic moment of the axialvector diquark and the 
vertex for photon--induced anomalous scalar--axialvector diquark 
transitions are transversal and need not be modified\cite{Oettel:2000jj}. 

\subsection{Strong Form Factors}
\label{strongFF}

Here we consider the strong form factors $g_{\pi NN}$ and
$g_{K N \Lambda}$. These quantities are not only interesting in
themselves but also enter the calculation of production processes like
$p\gamma \rightarrow \Lambda K$ or associated strangeness production
$pp \rightarrow pK\Lambda$.

\begin{figure}
\begin{center}
  \epsfig{file=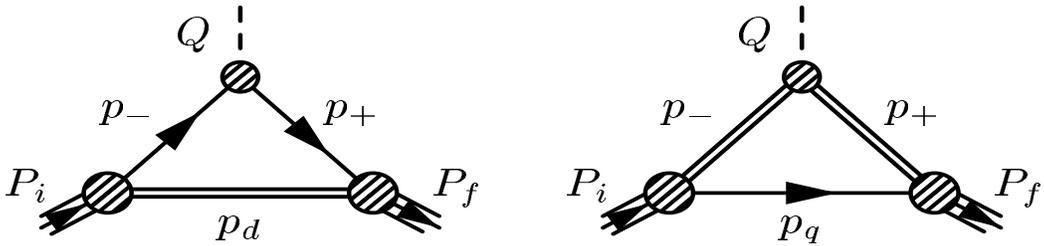,width=14.0cm,height=6.0cm}
\vskip-2cm
  \caption{\sf Dominant diagrams for the strong
           form factors $g_{\pi NN}$ and $g_{K N \Lambda}$.
           The incoming proton carries the momentum $P_i$ while $P_f$
           is associated with the outgoing baryon. The incoming
           meson carries the momentum $Q$ and couples to the
           quark (\emph{left panel}) or to the diquark (\emph{right panel}).}
  \label{sff_diag}
\end{center}
\end{figure}

In figure~\ref{sff_diag} we show the dominant contributions to the strong
form factors. Here the meson directly couples to one of the baryon
constituents. Keeping only such direct couplings while omitting those
to the exchanged quark defines the impulse approximation that we will
adopt here\footnote{Contributions to the nucleon electromagnetic form
factors beyond the impulse approximation that arise from the coupling
to the exchanged quark have been thoroughly discussed in
refs~\cite{Oettel:2000jj,Oettel:2000gc,Oettel:diss}.}.
The two diagrams shown
in figure~\ref{sff_diag} actually correspond to a number of diagrams
which differ by the type of the involved diquarks. Let us first
consider the process in which the meson couples to the quark. For
$g_{K N \Lambda}$ only one possibility exists: the diquark has
to be a scalar $ud$--diquark since this is the only overlap
between the wave--function of the proton and the wave--function of the
$\Lambda$. For $g_{\pi NN}$ both scalar and axialvector diquarks
need to be taken into account. The second important contribution represents
the coupling of the meson to the diquark. For the diquark part we do
not have to distinguish between $g_{\pi NN}$ and $g_{K N \Lambda}$.
That is, in both cases the diquark associated with the momenta $p_+$
or $p_-$ may be scalar or axialvector.

The meson--quark vertex is the solution of a separate Bethe--Salpeter
equation which has been extensively studied, see \cite{Alkofer:2000wg} and
references therein. In the chiral limit this Bethe--Salpeter equation
becomes formally identical to the Dyson--Schwinger equation for the
scalar self energy function $B(p^2)$ when only the leading Dirac structure
is considered, {\it i.e.}
\begin{equation}\label{qKVertex}
\begin{minipage}{2cm}
\epsfig{file=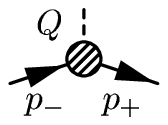}
\end{minipage}=
 \Gamma_m (p_-,p_+) = \frac{\imag}{2f} \gamma_5
\left\{B(p_+^2)+B(p_-^2)\right\}\, .
\end{equation}
Here $f$ is the meson decay constant.

The structure of the meson--diquark vertices is constrained not
only by Lorentz covariance and parity but also by the Bose--statistics
for the two involved diquarks. We thus parameterize the 
pseudoscalar meson axialvector diquark vertex as
\begin{equation}\label{piaa.vert}
  \Gamma^{\rho \lambda}_{aa} =
  -\imag\frac{\kappa_{aa}}{2M}
  \frac{m}{f} \epsilon^{\rho \lambda \mu \nu}
  (p_- + p_+)^{\mu} Q^{\nu}\, .
\end{equation}
Here the superscripts $\rho,\lambda$ denote the Lorentz indices
of the incoming and outgoing axialvector diquark, respectively. The
nucleon mass  $M$ has been introduced to define the dimensionless
coupling constant $\kappa_{aa}$. Furthermore $m$ is the average of
the masses of the constituent quarks in the diquarks. The 
corresponding {\it ansatz} for the scalar--axialvector transition reads,
\begin{equation}
\label{pisa.vert}
  \Gamma^{\rho}_{sa} =
  -\kappa_{sa} \frac{m}{f} Q^{\rho}\, ,
\end{equation}
where the definitions are those of eq~(\ref{piaa.vert}) and
$\kappa_{sa}$ is again a dimensionless constant specifying the
overall strength of the vertex. The vertex~(\ref{pisa.vert})
describes the coupling of the diquarks to the derivative of
the pseudoscalar mesons. Such a construction is suggested by
the chiral structure of the strong interactions that can
be written as expansion in the derivatives of the Goldstone
bosons, at least in the chiral limit.

Having collected all ingredients we may now proceed and compute the
diagrams in figure~\ref{sff_diag}. According to the Mandelstam formalism
\cite{Man55} the diagram shown in the left panel translates into an
expression
of the form
\begin{equation}\label{gknl_general}
  \int \frac{d^4q}{(2\pi)^4} \,
  \bar{\Phi}_f (q_f,P_f) S(p_+) \Gamma_{m} (p_-,p_+)
  S(p_-) \Phi_P(q,P_i)D(p_d)\, ,
\end{equation}
where we only indicated the general structure, {\it i.e.} we omit
indices that are associated to the coupling and propagation of
axialvector diquarks. The conjugated vertex function, $\bar{\Phi}$
relates to the vertex function, $\Phi$ as the conjugated wave
function~(\ref{psi_con}) to the wave function:
\begin{equation}
\bar{\Phi}(p,P)=\eta_d\, C\,\Phi(-p,-P)^T\, C^T
\label{phi_con}
\end{equation}
with $\eta_d=1$ and $\eta_d=-1$ when the involved diquark is
respectively of scalar or axialvector type. We note that
$\bar{\Phi}(p,P)$ also solves the Bethe--Salpeter equation~(\ref{BS2}).
We denote the loop momentum by $q$ and introduce the momentum partition,
\begin{equation}
   p_- = q + \eta P_i\, , \quad
   p_+ = p_- + Q = q_f +\eta P_f  \quad {\rm and} \quad
   p_d = -q  + (1-\eta) P_i\, .
\end{equation}
Again, $\eta \in [0,1]$ is the momentum partition parameter. For the
diagram in the right panel quark and diquark propagators need to be
exchanged.

\section{Bound State Reactions and Kinematical Conditions
for Complex Momenta}
\label{relevance}

In this section we will discuss that regime in the complex momentum 
plane where we need to know the quark and diquark propagators in order to solve
the Bethe--Salpeter equation~(\ref{bse_nuc}) and compare that regime to
the one that enters the computation of the production processes like 
$p\gamma \rightarrow \Lambda K$ and $pp\rightarrow pK\Lambda$. 
In principle these propagators can be calculated using Dyson--Schwinger
equations\cite{Alkofer:2000wg} and also respective lattice measurements 
should be available in the near future; for preliminary results 
see {\it e.g.} refs.~\cite{Skullerud:2000un,Hess:1998sd}. Both 
methods comprise the non--perturbative dynamics and should therefore 
give the basic ingredients to describe hadrons as
bound state of quarks. However, both approaches are set up in Euclidean 
space and one has to revert to extrapolations when the propagators are 
demanded for timelike momenta. If we wanted to perform an appropriate 
analytic continuation from Euclidean back to Minkowski space we would
even require the propagators in a region of the complex momentum
plane\footnote{In the Dyson--Schwinger approach the corresponding integral 
equation should be used for this analytic continuation. Relying on a
numerical solution that is only known for a finite set of Euclidean 
momenta is not sufficient because its analytic continuation away from 
that set cannot be determined.}.
In order to calculate amplitudes of physical processes between 
on--shell particles using the Euclidean Bethe--Salpeter formalism 
the temporal components of the external momenta must be purely imaginary. 
In this framework the momenta become complex. In these calculations
therefore the structure of the propagators in the 
complex momentum plane is essential. Furthermore, it is important for the 
phenomenological parameterization of confinement. As repeatedly mentioned 
we comprehend the confinement phenomenon as the absence of poles on the
timelike $q^2$--axis in the propagator of colored ``particles''. 

\begin{figure}
\centerline{
  \epsfig{file=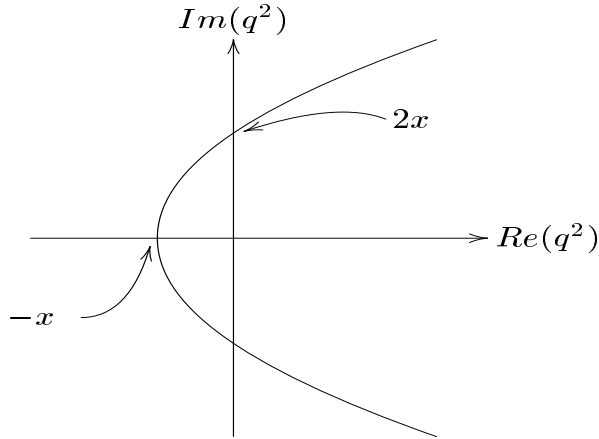,width=8cm,height=6cm}
}
\vskip0.5cm
  \caption{\sf The complex $q^2$--plane. The interior of the parabola 
     is needed for the calculation of the diagram in the left 
     panel of figure~\ref{kaon_hand_bag}, $x$ is
     defined as $x=(\eta M + E)^2$. Note that in the case of the 
     Bethe--Salpeter equation we substitute $x=\eta^2 M^2$.}
   \label{prod_kinem}
\end{figure}

The Bethe--Salpeter equation is most conveniently solved in the rest
frame of the bound state, $P=(\vec 0,iM)$. Here we want to specifically
discuss the kinematical domain that is probed by the (di)quark 
propagators in the bound state rest frame. In eq~(\ref{bse_nuc}) the
loop momentum, $k$, relative between quark and diquark is chosen
to be real. Hence the temporal component of the quark momentum $k_q=\eta P+k$
becomes complex. The values of $k_q^2$ that are covered when integrating
over $k$ lie within a parabola that opens towards the spacelike axis,
{\it cf.} figure~\ref{prod_kinem}. The intercept of the parabola with the
real axis is at (small) timelike $k_q^2=- (\eta M)^2$. Thus, solving the
Bethe--Salpeter equation mainly probes the behavior of the quark 
propagator for spacelike momenta. Since mainly the spacelike momenta are
relevant, the propagators that are parameterized by the dressing function 
(\ref{f2}) approach the free propagators in the limit 
$d\to+\infty$. In this limit the dressing functions (\ref{f1}) and 
(\ref{f3}) approach the bare propagators, both in the spacelike and 
timelike regions.
  
\begin{figure}
\centerline{
    \epsfig{file=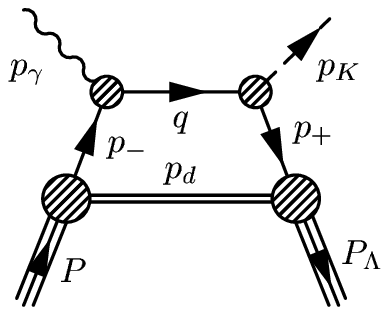,width=6cm,height=6cm}
\hspace{1cm}
    \epsfig{file=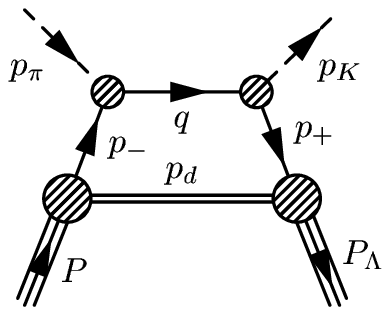,width=6cm,height=6cm}
}
\vskip0.5cm
    \caption{\sf {\it Left panel:} Main contribution to kaon
    photoproduction $p\gamma\to K\Lambda$, {\it right panel:} Handbag 
    diagram contributing to the reaction $pp \rightarrow pK\Lambda$ as 
    a subprocess. The incoming pion couples to the `spectating' proton.}
    \label{kaon_hand_bag}
\end{figure}

Next we will explore the $q^2$ regime relevant for production processes 
like kaon photoproduction. The contribution to the reaction 
$p\gamma\rightarrow\Lambda K$ that involves a quark loop
is shown in the left panel of figure~\ref{kaon_hand_bag}. It turns out 
that it suffices to consider a parabola shaped region of the complex 
$q^2$--plane ({\it i.e.} it is sufficient to consider only this 
momentum and ignore the others). This can be 
understood in at least two ways: We could use the wave--functions 
rather than the vertex--functions for the calculation of 
the diagram. In this case the propagators that depend on $p_q,k_q$ 
and $p_d$ are included in the wave--functions and there would be
no necessity to treat them separately. Nevertheless, considering the 
propagators $S(p_q),S(q),S(k_q)$ and $D(p_d)$ separately we find 
that among all the internal momenta in the diagram it is 
$q$ that reaches farthest in the timelike regime. Thus the following 
analysis for $p_q$ and $k_q$ would yield less restrictive conditions.
  
From the momentum routing shown in the left panel of 
figure~\ref{kaon_hand_bag} we have:
\begin{equation}
q = p_- + p_{\gamma} = \eta P + l + p_{\gamma}
\end{equation}
where $\eta$ is the momentum partition parameter
($p_-=\eta P+l,\; p_d=(1-\eta)P-l$) while $l$ refers 
to the loop momentum.  We chose the loop momentum
to be real which implies that the external momenta like $P$ and
$p_{\gamma}$ must have an imaginary temporal component in order to
correspond to physical particles.  
For the following kinematical
considerations we choose the proton rest frame and take the photon
to propagate along the $y$-axis,
\begin{equation}
  P = (\vec{0},\imag M),\quad
  p_{\gamma}=(0,E,0,\imag E),\quad
  l = (\vec{l},l_4\;)
\end{equation}
Hence the momentum entering the quark propagator becomes
\begin{equation}
q^2=\left(-\eta^2 M^2 + l^2 - 2\eta M E + 2 E l_y\right)
            + i \left( 2 \eta M + 2 E \right) l_4\, ,
\end{equation}
where the real and imaginary parts of $q^2$ have been separated.
This shows that we need to know the propagator $S(q^2)$ at 
complex $q^2$ in order to be able to compute the handbag 
diagram shown in the left panel of figure~\ref{kaon_hand_bag}.
The set of values of $q^2$ that occur has already been
shown in figure~\ref{prod_kinem}. The situation
seems to be completely parallel to what we found for the Bethe--Salpeter
equation; in both cases we need to know the propagators in a parabola 
shaped region of the complex plane. The intercept with the imaginary 
axis is in both cases minus two times the intercept with the real axis.
However, there is one important difference. For the production
processes the intercept with
the real axis does depend on the photon energy $E$, more
precisely: $-x=-(\eta M + E)^2$. Thus for $E=0$ the computation of the
handbag diagram shown in figure~\ref{kaon_hand_bag} uses the same
region of the complex plane that is necessary to solve the
Bethe--Salpeter equation. However, for $E>0$ the parabola is shifted in 
the direction of the negative real axis.
  
The threshold for kaon photoproduction is at $E$ slightly less than
$1{\rm GeV}$ and the cross section has been measured \cite{Tran:1998qw} 
up to $E\approx2{\rm GeV}$. This implies that the handbag diagram 
`probes' the quark propagator \emph{much} farther into the timelike 
region than the Bethe--Salpeter equation.

The second production process we are especially interested in is
associated strangeness production, $pp \rightarrow pK\Lambda$. This
reaction can be described similarly to the standard picture of
the nucleon--nucleon interaction by one--boson exchange. 
That is, one of the incoming protons acts as a meson source 
and the emitted off--shell meson couples to one of the constituents of 
the baryon; the corresponding subprocess is shown diagrammatically 
in figure~\ref{kaon_hand_bag}.

The analogous kinematical analysis for strangeness--production exhibits 
the same qualitative features.  That process as well `probes' a 
parabola shaped subset of the complex plane, whereby the parabola is 
somewhat broader than the one in figure~\ref{prod_kinem}. However, 
there again is an important difference: the parabola does extend only 
up to $q^2 \approx-0.53$GeV$^2$ into the timelike region. That is, the 
reaction $pp \rightarrow pK\Lambda$ `probes' the propagators in essentially 
the same region as the Bethe--Salpeter equation does. It is therefore not 
as sensitive as kaon photoproduction to the behavior of the propagator 
in the timelike region.

The main conclusion of the above discussion is that certain
production processes may be significantly more sensitive to
the structure of the (di)quark propagators than the Bethe--Salpeter
equation and thus the baryon spectrum. Hence the study of such 
processes should provide important information about these propagators.
 
\section{Production processes in the diquark--quark model}
\label{CalcObserv}

In this section we present the key issues of the formalism to 
compute the cross sections for 
kaon photoproduction, $\gamma p \rightarrow \Lambda K$ and the associated 
strangeness production, $pp \rightarrow pK\Lambda$. In the 
diquark--quark model relatively few diagrams contribute to these 
processes and therefore we may analyze these reactions in detail.
For further details on the definition of the involved observables and 
the relevant kinematics we refer the reader to appendix~\ref{app_product}.

\begin{figure}[t]
\begin{center}
  \epsfig{file=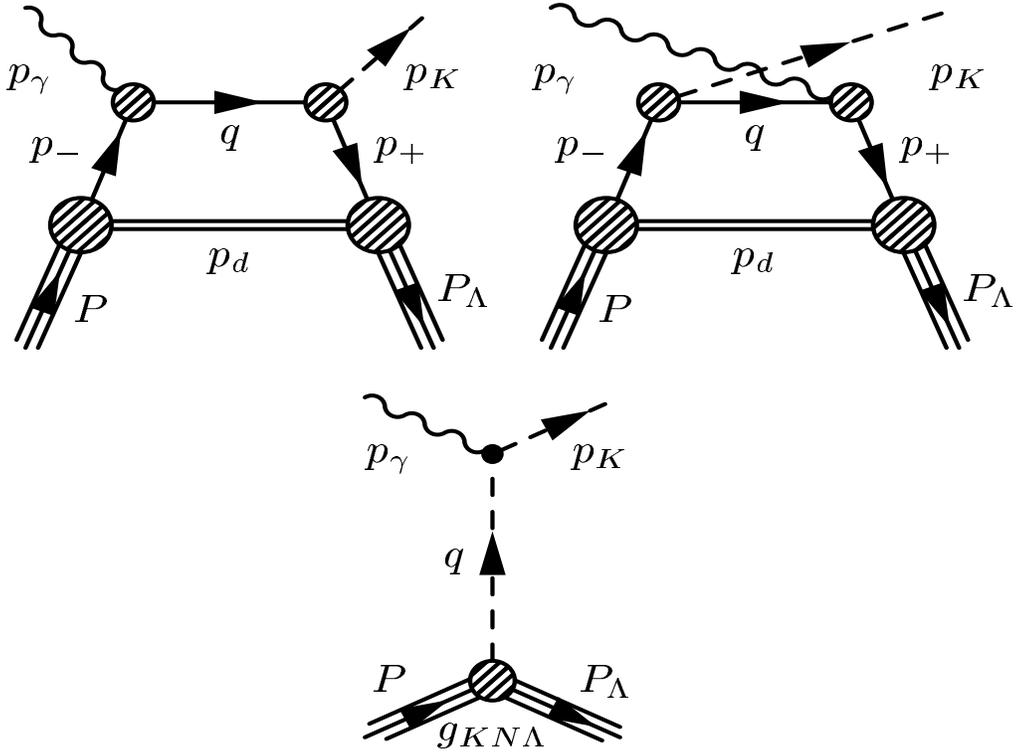,width=14.0cm,height=10.0cm}
\vskip1.0cm
  \caption{\sf Main contributions to kaon photoproduction
     $p\gamma \rightarrow \Lambda K$. The incoming proton and
     the outgoing $\Lambda$ carry the momenta $P$ and $P_\Lambda$
     respectively. The lower part of the figure shows the 
     tree level diagram that models the exchange of a virtual kaon.}
  \label{kp_photo_diag}
\end{center}
\end{figure}

As already indicated in the discussion of the strong form factors
we consider the pseudoscalar mesons as additional model degrees of
freedom. This does not imply any double counting because the model 
interaction (diquark exchange) does not lead to bound (would--be) 
Goldstone bosons. Thus we also include intermediate pseudoscalar mesons 
at tree level when computing the above mentioned observables. The 
relevant diagrams are shown in figures~\ref{kp_photo_diag} 
and \ref{pppkldiag}. As a general remark we note that these diagrams 
need to be computed in any covariant diquark--quark model. However, 
the propagators that are essential components of these diagrams are 
specific to our model, {\it cf.} eqs~(\ref{f0})--(\ref{f3}).
Furthermore, the covariant wave-- or vertex--functions that also enter
these diagrams are obtained as solutions of the Bethe--Salpeter 
equation. Since this equation is subject to the model propagators they
enter the calculation not only explicitly but also implicitly.

\subsection{Kaon Photoproduction}\label{KaonPhoto}

In this subsection we will discuss kaon photoproduction $p\gamma
\rightarrow \Lambda K$ within our covariant diquark--quark model.  
Some more details and technicalities of the calculation are given 
in appendix \ref{app_KaPho}.

We show the dominant diagrams in figure~\ref{kp_photo_diag}. 
The internal momenta of the (uncrossed) `handbag diagram'
are defined according to
\begin{eqnarray}
  p_- &=& p + \eta_p P \, ,\quad
  p_d = -p + (1-\eta_p ) P  \, ,\quad\
  q   = p_- + p_{\gamma}\, , \nonumber    \\
  p_+ &=& q - p_{K}  \, ,\quad
  P_\Lambda = P + p_{\gamma} - p_K  \, ,\quad
  p_f = p + (1-\eta_p ) P - (1-\eta_{\Lambda} ) P_\Lambda \, .
\label{A1.mom.1}
\end{eqnarray}
Here $\eta_p$ and $\eta_{\Lambda}$ are the momentum partition
parameters of the proton and the $\Lambda$, respectively. Both,
$\eta_p$ and $\eta_{\Lambda}$ can be chosen independently in the 
range $0 \le \eta_p,\eta_\Lambda \le 1$.

The two `handbag diagrams' model the coupling to one of the constituents.
They are calculated within the Mandelstam formalism.
This yields
\begin{equation}
A_1=\imag\int\frac{d^4 p}{(2\pi)^4}
\left\{\bar{\Phi}_{\Lambda}(p_f,P_\Lambda)S(p_+)
\Gamma_K(q,p_+)S(q)\right\}
\left\{\Gamma_{\gamma}(p_-,q) S(p_-) \Phi_P(p,P) D(p_d)\right\}
\end{equation}
for the amplitude of the uncrossed handbag diagram. Here
$\bar{\Phi}_{\Lambda}$ and $\Phi_p$ are respectively the 
vertex--functions of the $\Lambda$ and the proton as discussed in 
section~\ref{dqmodel}. Furthermore $\Gamma_K$ is the meson--quark
vertex that has been discussed in the preceding subsection.
The photon--quark coupling, $\Gamma_{\gamma}$ is described by the
Ball--Chiu vertex~\cite{Ball:1980ay} or its generalization to the case of
non--analytic propagators, see section~\ref{dqmodel}. The Ball--Chiu
vertex has been constructed to satisfy the Ward identity. It reduces to 
the bare vertex in the limit that both momenta $p$ and $q$ are large. 
The Ward identity constrains only the longitudinal part of the vertex. 
Various {\it ans\"atze} for the transversal part of the vertex have been 
proposed ({\it cf.} ref~\cite{Alkofer:2000wg} and references therein). 
While those {\it ans\"atze} solve problems related to multiplicative 
renormalizability and gauge invariance, the transversal part is generally 
assumed to be of minor influence on the resulting cross sections. Thus we 
will henceforth neglect the transversal part of the quark--photon vertex. 
Although the form of this vertex is not model specific, it contains the
self--energy functions and thus it implicitly depends on the model
propagators. The expression for the crossed handbag diagram can be easily 
inferred. The tree level diagram models the exchange of a virtual kaon 
and is expected to yield a non--negligible contribution for large photon 
energies. For the photon--meson coupling we use a bare vertex multiplied
with the kaon electromagnetic form factor (see Appendix \ref{app_KaPho}) while 
the  meson--baryon vertex is proportional to $g_{KN\Lambda}(Q^2)$ that has 
been discussed in subsection~\ref{strongFF}. 

The `handbag diagrams' shown in figure~\ref{kp_photo_diag} probe the 
propagators not only for spacelike momenta but also for comparably large 
timelike momenta, as we have emphasized in section~\ref{relevance}. 
This sensitivity to the behavior of the propagators for timelike 
momenta distinguishes the reaction $p\gamma\rightarrow\Lambda K$ from 
most other production processes.

\subsection{Associated Strangeness Production}

\begin{figure}[tp]
\centerline{
  \epsfig{file=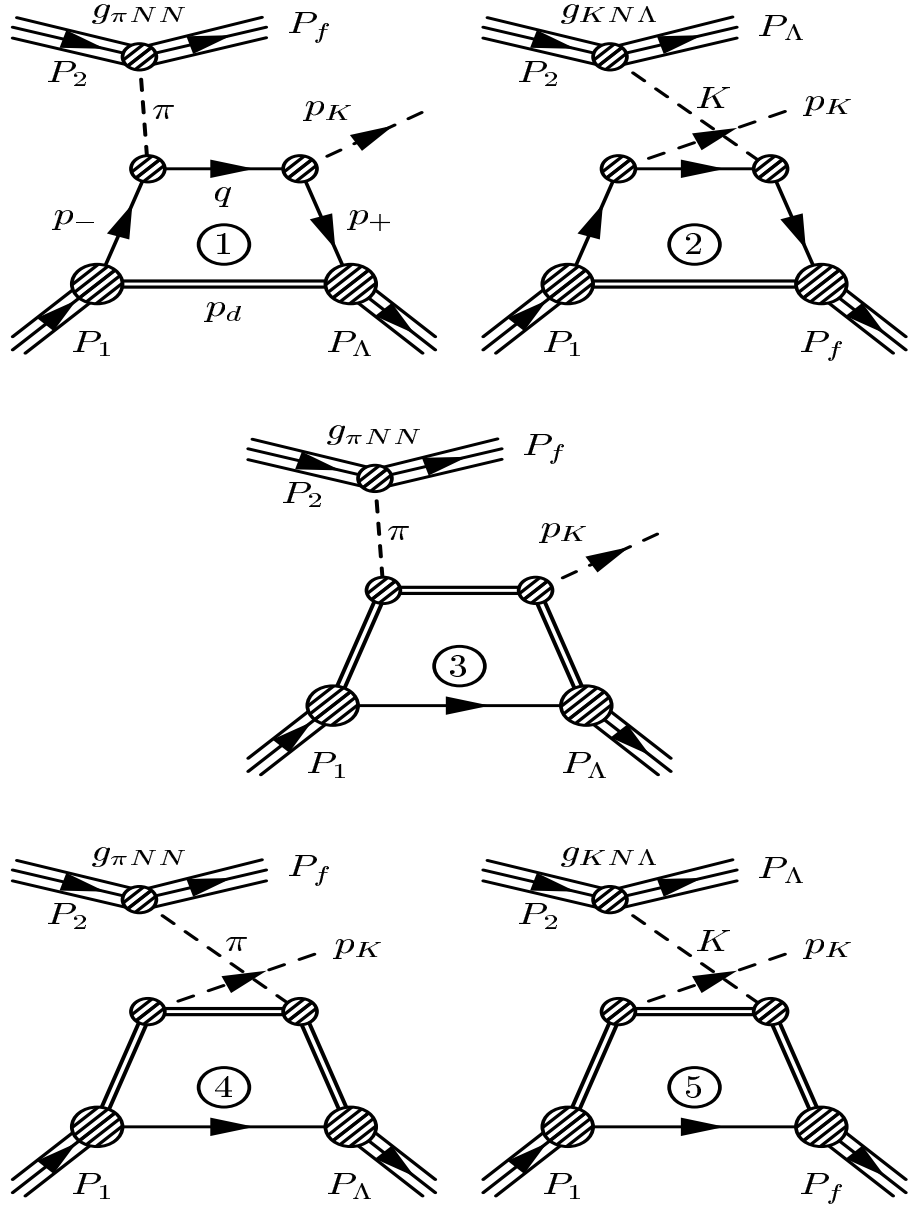,width=12.0cm,height=16.0cm}
}
\vskip1.0cm
\caption{\sf  Main contributions to associated strangeness production
          $pp \rightarrow pK\Lambda$. The momenta of the initial protons
          are given by $P_i\, , i=1,2$. The momenta of the final proton
          and $\Lambda$ are denoted by $P_f$ and $P_\Lambda$, respectively.
          $p_K$ refers to the momentum of the final kaon. The diquark 
          content of the diagrams is listed in Table \ref{tab_AsStr_1}.}
\label{pppkldiag} 
\end{figure}

Here we apply the covariant diquark--quark model to associated 
strangeness production, $pp \rightarrow pK\Lambda$. Again, some 
technicalities are relegated to appendix~\ref{ap_ASPRO}. We describe 
the reaction $pp \rightarrow pK\Lambda$ as a sum over one--boson 
exchange contributions for which we consider the exchange of pions 
and kaons. The main contributions are shown in figure~\ref{pppkldiag}. 
Other diagrams like planar kaon exchange or crossed pion exchange with 
couplings to the quark are excluded by the flavor algebra. We neglect 
diagrams in which the exchanged particle and the emitted kaon couple 
to different constituents, because they imply a large relative momentum 
at the baryon vertex which is strongly suppressed. In addition we omit 
the direct coupling of the pion to the emitted kaon.

\begin{table}
\centering
\begin{tabular}{l|cccc}
\textit{diagram \#} & 1 & 2 & 3,5 & 4 \\
\textit{diquark content} 
& s & s,a & asa, aas, aaa, sas, saa & sas, saa \\
\end{tabular}
\caption{\sf The diquark content of the diagrams shown in 
figure~\ref{pppkldiag} with the numbers referring to the specific
diagram. Here 's' and 'a' indicate scalar and axialvector diquark 
respectively. A sequence with three entries runs clockwise and starts 
at the incoming proton.}
\label{tab_AsStr_1}
\end{table}

Similar to the calculation of $g_{KN\Lambda}$ in section~\ref{strongFF},
the flavor content of the $\Lambda$ prohibits an axialvector diquark in 
the pion exchange diagram \#1. However, this does not apply to the kaon 
exchange diagram \#2 as the diquark mediates between two protons. Due to 
parity conservation we only have to consider scalar--axialvector and 
axialvector--axialvector diquark transitions at any meson--diquark vertex. 
Flavor algebra also shows that the axialvector diquark components of the
incoming proton cannot contribute in diagram \#4. The diquark content
of the diagrams shown in figure~\ref{pppkldiag} is listed in
table~\ref{tab_AsStr_1}. Counting the diquark combinations in
table~\ref{tab_AsStr_1} we arrive at 15 diagrams. This number is
actually doubled because all diagrams have to be antisymmetrized with
respect to the two incoming protons.

As an example, we outline the calculation of one of the two planar 
pion exchange diagrams. The calculation of the other diagrams is 
very similar. The amplitude of diagram \#1 can be factorized
according to
\begin{equation}
{\mathcal M} = {\mathcal L} \, 
\left( \frac{i}{Q^2+m_{\pi}^2} \right) \, {\mathcal H}\, ,
\label{pidia1}
\end{equation}
where  ${\mathcal L}$ denotes the form factor part and ${\mathcal H}$ 
denotes the loop part of the diagram. The mass and the momentum of the 
intermediate pion are denoted by $m_{\pi}$ and $Q$, respectively.
The factor ${\mathcal L}$ essentially equals $g_{\pi NN}$,
\begin{equation}  
{\cal L}_{s,s'} = \bar{u}_{s'}(P_f) \: i\gamma_5 \, 
g_{\pi NN}(Q^2)\: u_s(P_2),
\label{pidia2}
\end{equation}
with spinor indices $s$ and $s^\prime$. In the remaining `handbag part' 
${\mathcal H}$ the conventions for the loop momenta can be extracted 
from diagram \#1 of figure~\ref{pppkldiag}. Essentially they 
are given in eq~(\ref{A1.mom.1}) with the substitution 
$p_\gamma\to Q$ and similarly for the baryons.
The `handbag part' of the pion exchange diagram can then be written as 
\begin{equation}\label{pidia3}
{\mathcal H}=i\int\frac{d^4p}{(2\pi)^4}\left\{
\bar{\Phi}_{\Lambda}(p_f,P_\Lambda)\, S(p_+)\, 
\Gamma_K(p_+,q)\,S(q)\right\}
\left\{\Gamma_\pi(q, p_-)\,S(p_-)D(p_d)\,
\Phi_{P}(p_1,P_1)\right\}\, ,
\end{equation}
where isospin as well as Lorentz indices have been omitted for
simplicity.

\section{Numerical Results}
\label{num_res}
After having outlined the model calculation we are now
prepared to present the numerical results. Here we focus on studying 
the effects of the different model 
propagators~(\ref{f0})--(\ref{f3}) on the calculation of and
the predictions for the above mentioned processes.\footnote{A few selected 
numerical results using other forms of dressed model propagators have been published 
in refs.\ \cite{Alk00}.} As mentioned
earlier, this is the main purpose of the present study. 
Numerical results for the form factors
obtained with the tree level propagators (\ref{f0}) can be 
found in ref.~\cite{Oettel:2000jj}. 

\subsection{Parameter Fixing: Masses and Electromagnetic Form Factors}
\label{para_fix}

We fix the model parameters, see table~\ref{para_tab}, from the octet 
baryon masses and the nucleon magnetic moments. The numerical details 
for solving the octet baryon Bethe--Salpeter equations and the computation 
of the form factors are thoroughly discussed in ref.~\cite{Oettel:diss}. 

Within the required 
numerical accuracy we have assured the above described independence of the 
octet masses of the momentum partition parameter $\eta$ when analytical
propagators are used. As argued before, this invariance does not hold 
for non--analytic propagators. In these cases we chose $\eta$ to 
be close to its non--relativistic value $m_q/(m_q+m_d)$ where $m_q$ and 
$m_d$ denote quark and diquark mass parameters of the flavor channel 
associated with the considered baryon. This choice is natural since 
other ones yield larger eigenvalues of the Bethe--Salpeter equation.
We take the physical nucleon mass to fix the scalar diquark coupling $g_s$
and the $\Lambda$ mass to determine the strange quark mass parameter $m_s$.
By reproducing the phenomenological dipole fit for the proton electric form
factor, $G_E$ we essentially fix the diquark width $\lambda$. Subsequently
we are enabled to compute the proton and neutron magnetic moments, $\mu_p$
and $\mu_n$ as well as the masses of the $\Sigma$ and $\Xi$
baryons. For that calculation we assume isospin symmetry, $m_u=m_d$.

In table \ref{para_tab} we list the six parameter sets that we will
employ to compute the strong form factors and observables of production 
processes later on. The first four sets are restricted to the dominant 
scalar diquark correlations. In set I we consider the pole--free exponential 
dressing function, $f_2$, while the sets II and III are associated with 
dressing functions of the Stingl type, $f_1$. These two sets differ 
by the value of $d$ that characterizes the separation of the complex
conjugated poles. Finally set IV assumes the non--analytic pole--free 
dressing function, $f_3$. As already indicated the dressing of the 
propagators increase the predicted proton magnetic moment when all 
other model parameters remain unchanged. Using the parameters of set II
but free propagators yields $\mu_p=2.27$ while the Stingl--type 
propagators result in $\mu_p=2.46$ and $\mu_p=2.64$ for $d=8.0$ and 
$d=4.0$, respectively. The magnetic moment of the proton falls a 
little short for the sets II and IV. The overall picture emerges that 
the restriction to only scalar diquark correlations produces too 
large ratios $|\mu_n/\mu_p|$ and rather large mass splittings
between the octet baryons, especially between $\Sigma$ and $\Lambda$.

\begin{figure}[t]
\centerline{
 \epsfig{file=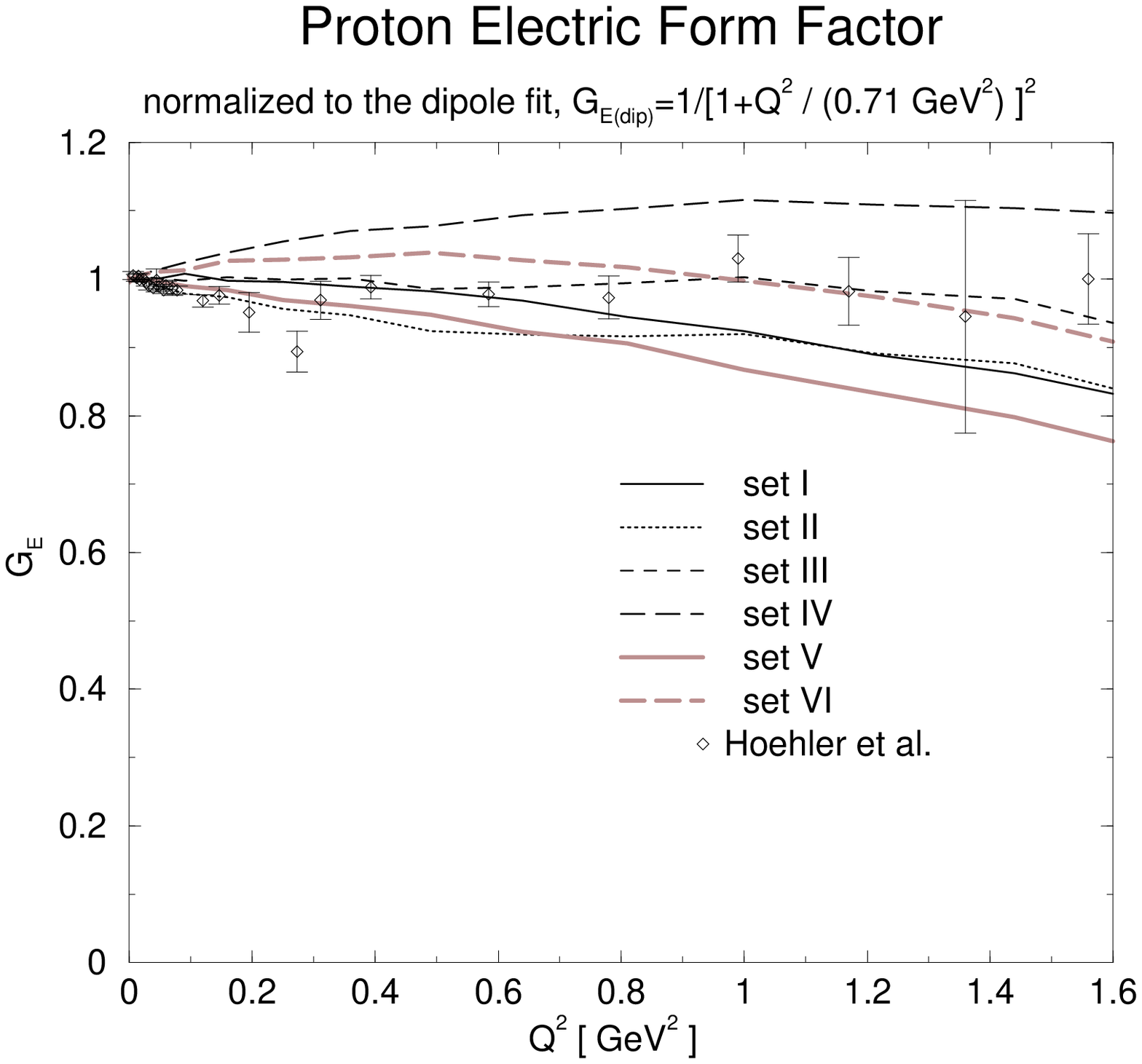,width=7cm}
 \hspace{1cm}
 \epsfig{file=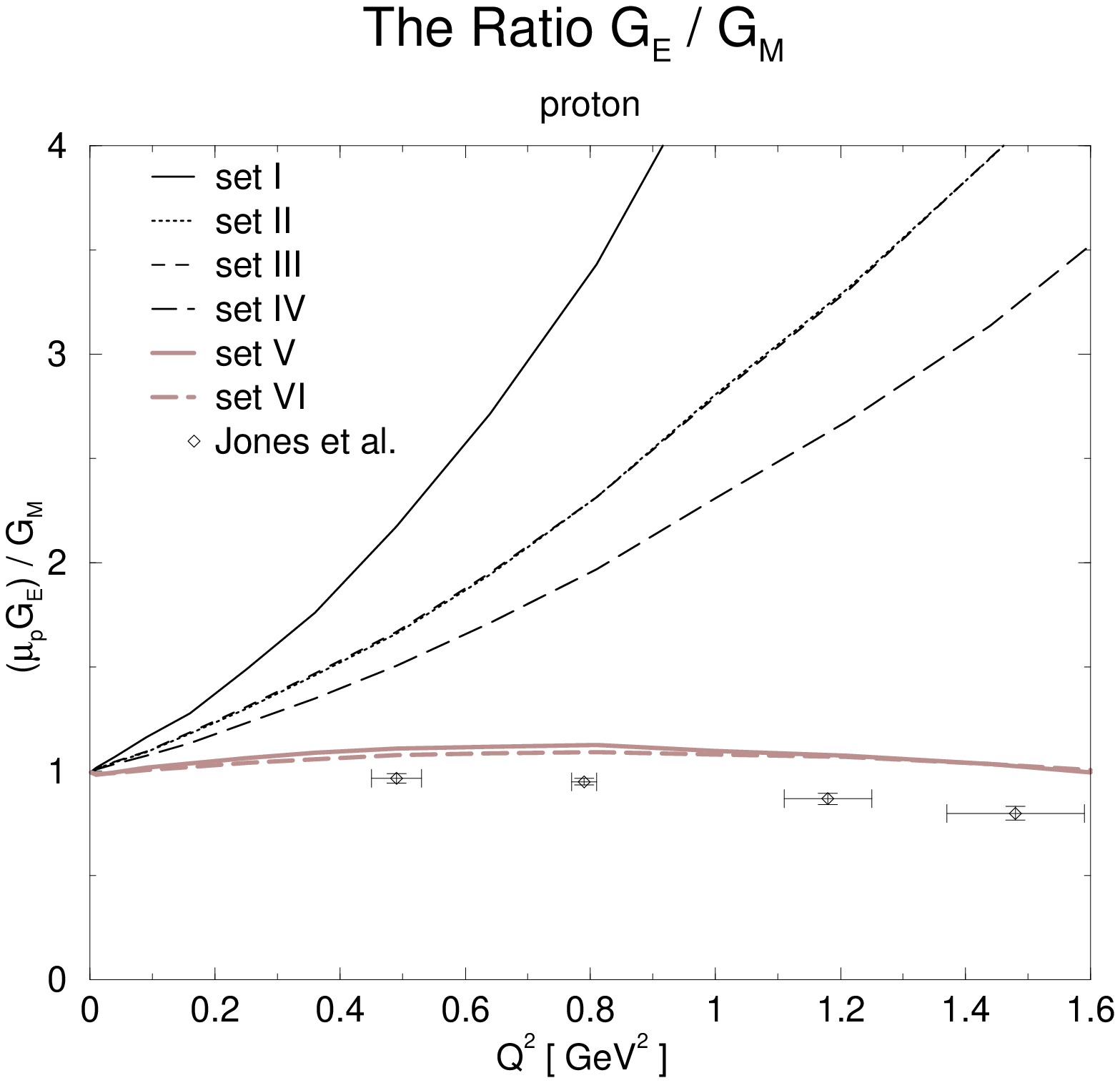,width=7cm}
}
 \caption{\sf \label{ge}
 Left panel: Proton electric form factor normalized to the dipole 
 fit. The experimental data are taken from ref.~\protect\cite{Hohler:1976ax}.
 Right panel: The ratio $(\mu_p G_E)/G_M$ for the proton with the
 experimental data published in ref.~\protect\cite{Jones:2000rz}.}
\end{figure}

As shown in figure~\ref{ge} all sets reasonably well reproduce the 
electric form factor, $G_E$. Our results are confined within a region 
that is characterized by less than approximately 15\% deviation from the 
dipole fit. This deviation, although rectifiable by refining the 
time--consuming parameter search, is of no significance for the 
conclusions that we will draw from our results for the production 
processes. This will become clear from the discussions in 
section~\ref{prod_proc}.

\renewcommand{\arraystretch}{0.7}
\begin{table}[t]
\begin{tabular}{rrrrrrrr}
 & I & II & III & IV & V & VI & expt. \\
 diquark:
 & \multicolumn{4}{c}{only scalar} & \multicolumn{2}{c}{scalar and}\\
 & & & & & \multicolumn{2}{c}{axialvector} \\ 
 $f_i$ & 2 & 1 & 1 & 3 & 1 & 3 \\
 $d$          & 2.0 & 8.0  & 4.0 & 6.0 & 4.0  & 6.0 \\
 $m_u=m_d$[GeV] & 0.40 & 0.45 &0.45 & 0.52 & 0.45 & 0.52 \\
 $m_s$ [GeV] & 0.64 & 0.70 & 0.69 & 0.75 & 0.67 & 0.72 \\
 $\zeta$ & 0.70 & 0.95 & 0.92 & 0.97 & 0.92 & 0.97 \\
 $\lambda^2$ [GeV$^2$] & 0.25 & 0.1 & 0.1 & 0.1 & 0.1 & 0.1 \\ \hline \\
 $\mu_p$ & 2.83 & 2.47 & 2.64 & 2.32 & 2.70 & 2.33 & 2.79 \\
 $\mu_n$ & $-$2.37 & $-$2.15 & $-$2.32 & $-$2.08 &  
  $-$2.08 &$-$1.82& $-$1.91  \\ \hline \\
 \multicolumn{7}{c}{octet masses, $M_N=0.939$ GeV fixed}\\
 $\Lambda$  [GeV] & 1.13 & 1.12 & 1.12 & 1.12 & 1.13 & 1.12 & 1.12\\
 $\Sigma$   [GeV] & 1.30 & 1.27 & 1.29 & 1.30 & 1.22 & 1.21 & 1.19\\
 $\Xi$      [GeV] & 1.37 & 1.37 & 1.39 & 1.36 & 1.37 & 1.33 & 1.32 
\end{tabular}
\caption{\sf \label{para_tab}
The six parameter sets of the model investigated here 
and the respective results for the nucleon magnetic moments and the 
octet masses. Calculations using the first four sets involve only 
scalar diquarks, whereas the sets V and VI also include axialvector 
diquarks.  The parameter $\zeta$ determines the diquark mass parameter
(scalar and axialvector), $m_d=\zeta(m_a+m_b)$, with the mass parameters 
$m_{a,b}$ of its constituent quarks. The parameter $\lambda$ determines 
the width of the diquark amplitudes, see eq~(\ref{npole}). For set I 
the corresponding shape of the amplitudes was chosen to be a quadrupole 
($n=4$), for the other sets we fixed it to be a dipole ($n=2$). }
\end{table}

The calculations with the parameters sets V and VI include a moderate
admixture of axialvector diquarks, $g_a/g_s=0.2$. For simplicity the 
axialvector diquark masses are chosen identical to the scalar ones.
Here we particularly
consider the Stingl form, $f_1$ (set V) and the non-analytic form, $f_3$ 
(set VI) since later we will find that the exponential form, $f_2$ produces 
unacceptable results for the production processes. Upon inclusion of the 
axialvector diquark the good description of $G_E$ remains unchanged while 
the ratio $|\mu_n/\mu_p|$ and the mass splitting between $\Sigma$ and 
$\Lambda$ even improve. For set VI the predicted octet masses are almost 
indistinguishable from their experimental values. As already observed in 
ref.~\cite{Oettel:2000jj} and as is exhibited in the right panel of 
figure~\ref{ge}, the ratio $G_E/G_M$ calculated with axialvector diquarks 
included comes considerably closer to the experimental values than in a
calculation that omits these degrees of freedom (sets I--IV). As explained 
in ref.~\cite{Oettel:2000jj}, increasing the strength of axialvector 
correlations in the proton forces the ratio $G_E/G_M$ to bend to lower 
values. This also suggests that in order to precisely reproduce the 
empirical result we would need an even slightly larger axialvector 
coupling than the assumed $g_a/g_s=0.2$.

All sets predict the maximum of the neutron electric form factor to lie
between 0.025 and 0.04. This is only about half the value extracted from
recent experiments~\cite{Passchier:1999cj,Ostrick:1999xa}. This form 
factor is a result of delicate cancellations between the contributions 
from the individual diagrams shown in figure~\ref{emff}. Hence it is 
quite sensitive to small changes in the parameters. Within this model
approach improved descriptions for this form factor can be found in 
refs.~\cite{Oettel:2000gc,Oettel:2000jj}. 

In a previous study \cite{Oettel:2000jj} that employed free quark and
diquark propagators we were unable to reproduce the nucleon magnetic
moments and the $\Delta$ mass simultaneously. The kinematical binding
of the $\Delta$ required a large constituent quark mass,
$m_q=0.43$ GeV, which in turn decreased the magnetic moments (in
magnitude). Furthermore the use of free propagators enforced moderate
axialvector diquark contributions (about 25\%) to properly describe the
ratio $G_E/G_M$ of electric and magnetic form factors for $Q^2$ up to
$2{\rm GeV}^2$. In contrast, the introduction of dressing functions for
the quark--photon vertex~(\ref{wtf012},\ref{wtf3}) allows us  to
choose rather large {\em up} quark mass parameters around $m_u=0.45$ GeV
and still obtain a proton magnetic moment that agrees with 
experiment reasonably well.

Let us briefly reflect on the accuracy of our calculations.
Due to the Monte--Carlo integration of the diagrams given in part $b)$ of
figure~\ref{emff} (with $7.5 \cdot 10^5$ grid points for sets I-IV and
$4 \cdot 10^5$ grid points for sets V and VI) the absolute numerical error
for $\mu_p$ is 0.02 and for $\mu_n$ it is 0.03. The statistical error for
the electric form factor is below 0.002 up to momentum transfers of 
$1.7{\rm GeV}^2$. For the sets IV and VI a 
systematic relative error in the electromagnetic form factors is found
that increases slowly to about 5 \% at $Q^2=1.7{\rm GeV}^2$.
Furthermore, we used an expansion in Chebyshev polynomials for wave and 
vertex functions because this expansion can be unambiguously continued 
to complex arguments. Upon employing analytical propagators we have 
obtained identical results when using the approaches wherein either
the wave functions $\Psi$ or the vertex functions $\Phi$ must be 
continued analytically~\cite{Oettel:2000gc}. For non--analytic
propagators, the corresponding solutions of the Bethe--Salpeter equation
are non--analytic as well. Thus the analytic continuation produces errors
which can be estimated by comparing the calculations using wave functions 
or vertex functions. In order to restrict the resulting 
discrepancy to below 5\%, we have to choose $d>5$ and rather large quark and 
diquark mass parameters. This then leads to nucleon solutions of the
Bethe--Salpeter equation that resemble an analytic function in the 
kinematic domain needed for the ongoing computation.

\subsection{Strong Form Factors}
\label{strong_ff}

Above we have utilized baryon properties to determine the model 
parameters that enter the Bethe--Salpeter problem. Furthermore the
meson--quark vertex is governed by the appropriate Ward--Takahashi 
identity. Finally the 
meson--diquark coupling constants $\kappa_{sa}$ and $\kappa_{aa}$ in 
eqs~(\ref{piaa.vert}) and~(\ref{pisa.vert}) have already been 
determined in ref.~\cite{Oettel:2000jj}. Thus we are now completely
prepared to compute the loop--integrals like that in 
eq~(\ref{gknl_general}) and that appear in figure~\ref{sff_diag}.  
Subsequently we may extract $g_{\pi NN}$ and $g_{KN\Lambda}$. 

There have been numerous experimental efforts to determine the strong 
form factors. An extended discussion of the phenomenological value 
of $g_{\pi NN}$ and a comprehensive list of related references is 
provided in ref.~\cite{Ericson:2000md}. Mainly the quoted discrepancies 
are subject to different analysis of available data. For the purpose of 
the present work it is sufficient to know that the quoted data are of the 
order $g_{\pi NN} \approx 14$. Unfortunately the measurements 
of $g_{KN\Lambda}$ have not yet reached a satisfactory accuracy. The 
authors of~\cite{Timmermans:1995qa} have extracted 
$|g_{KN\Lambda}(Q^2 = -M_K^2)|=13.7 \pm 0.9$ from the LEAR--data; but 
other analysis have partially yielded quite different 
results~\cite{Adelseck:1990ch,Adelseck:1992ua}.

\begin{figure}
\centerline{
\epsfig{file=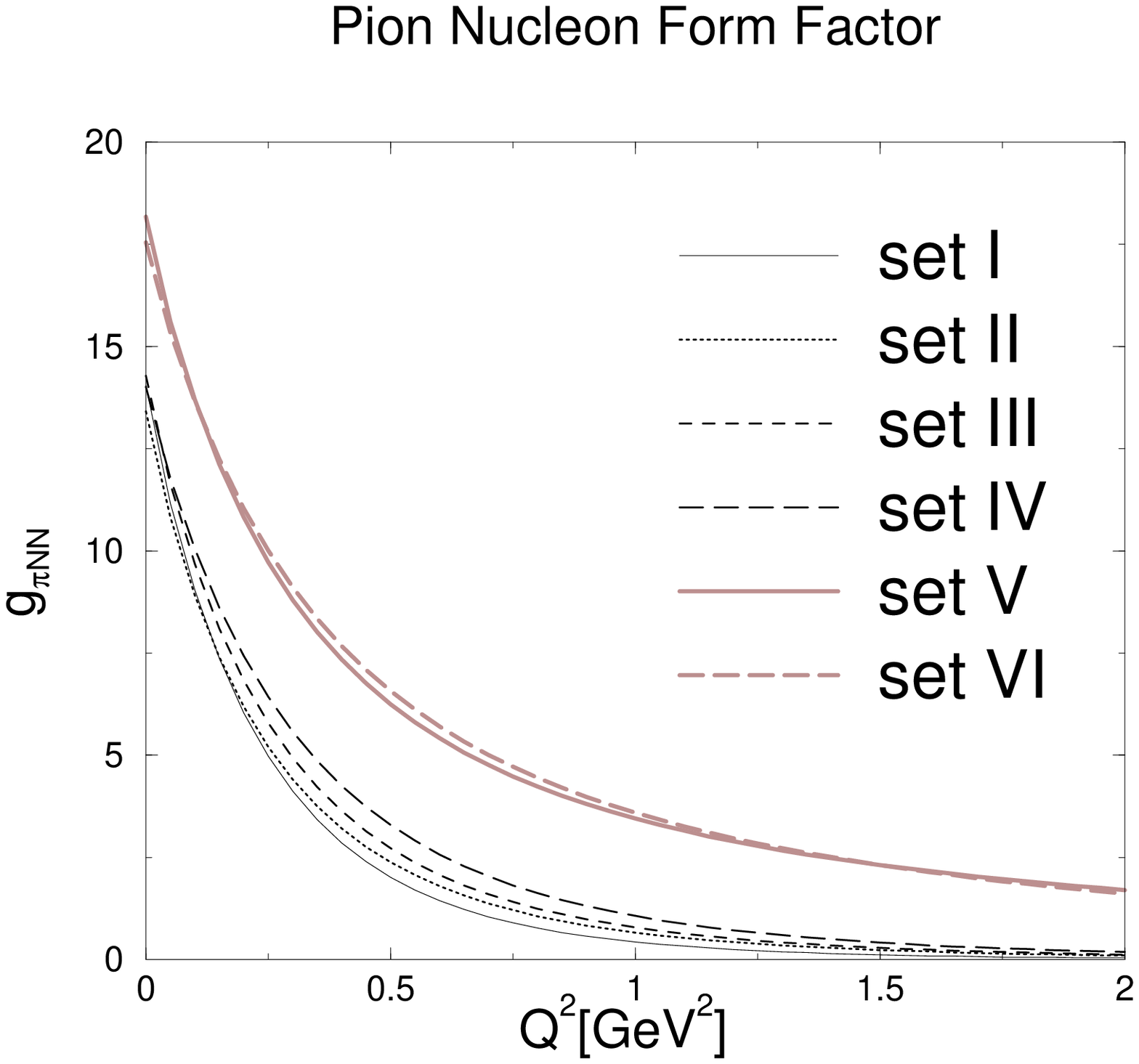,width=6.0cm,height=6.0cm}
\hspace{1cm}
\epsfig{file=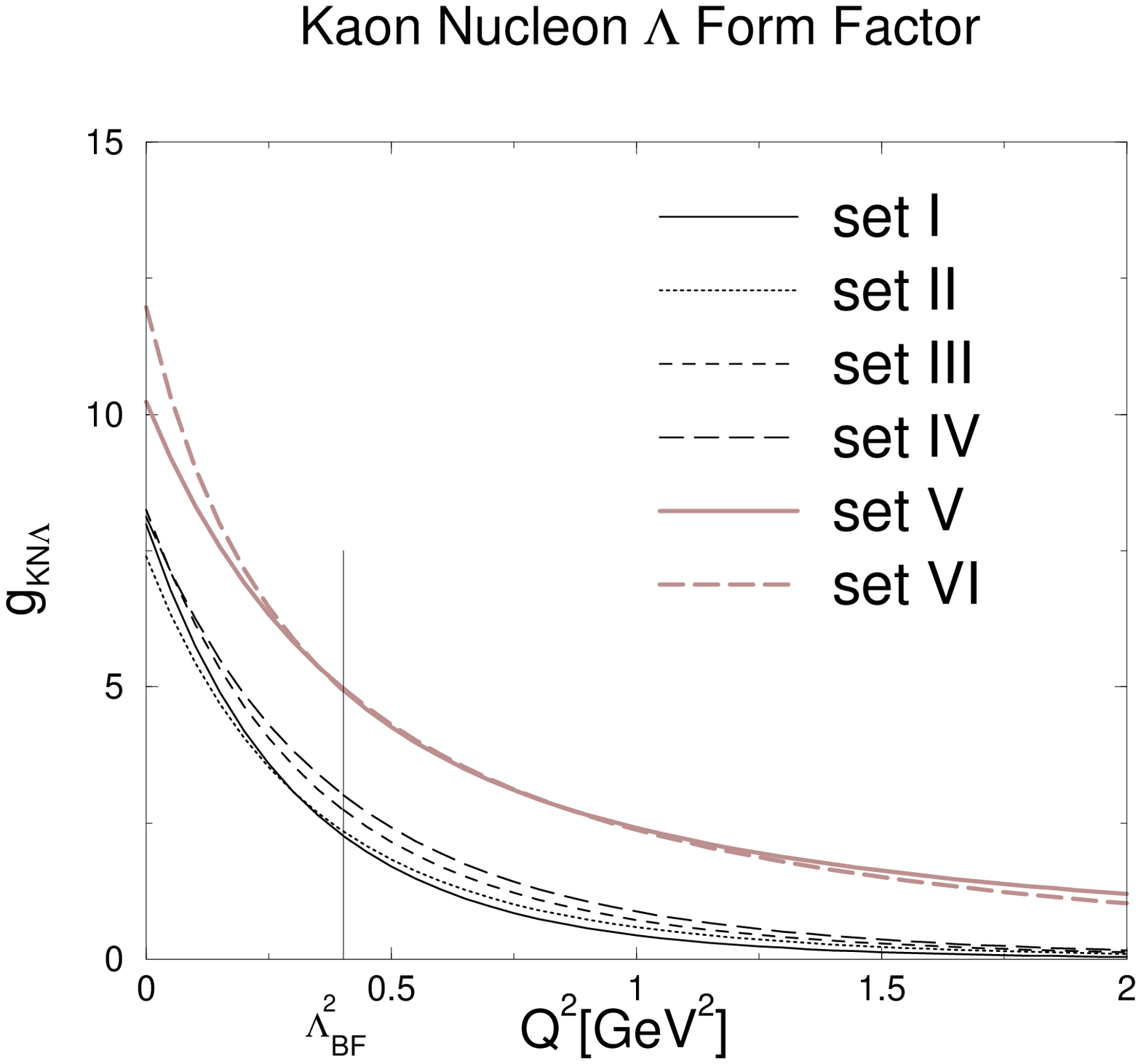,width=6.0cm,height=6.0cm}
}
\caption{\sf \label{g_all}
The model prediction for the strong form factors $g_{\pi NN}$ and 
$g_{KN\Lambda}$. The parameter sets (I--VI) are defined in 
table~\protect\ref{para_tab}. For $g_{KN\Lambda}$ the results 
in the region $Q^2<\Lambda^2_{\rm BF}$ are obtained from
a rational fit (see text).} 
\end{figure}

In figure~\ref{g_all} we display the numerical results for the 
$g_{\pi NN}(Q^2)$ and $g_{KN\Lambda}(Q^2)$. Both form
factors have been calculated in the Breit--frame. This frame is peculiar 
because for different masses of the initial and final baryons a
numerically save treatment induces a lower bound 
($Q^2\ge\Lambda^2_{\rm BF}=M_\Lambda^2-M_N^2\approx0.4{\rm GeV}^2$) 
for the momentum, $Q$, of the coupling meson.
In the case of $g_{KN\Lambda}$ we have fitted the computed
form factor to rational functions and extrapolated those functions
to $Q^2\to0$. The resulting coupling constants are shown in 
table~\ref{form_tab}. For the special case $Q^2=0$ we have verified 
that this treatment yields the same result as a calculation in 
the lab--frame.

\begin{figure}
\centerline{
\epsfig{file=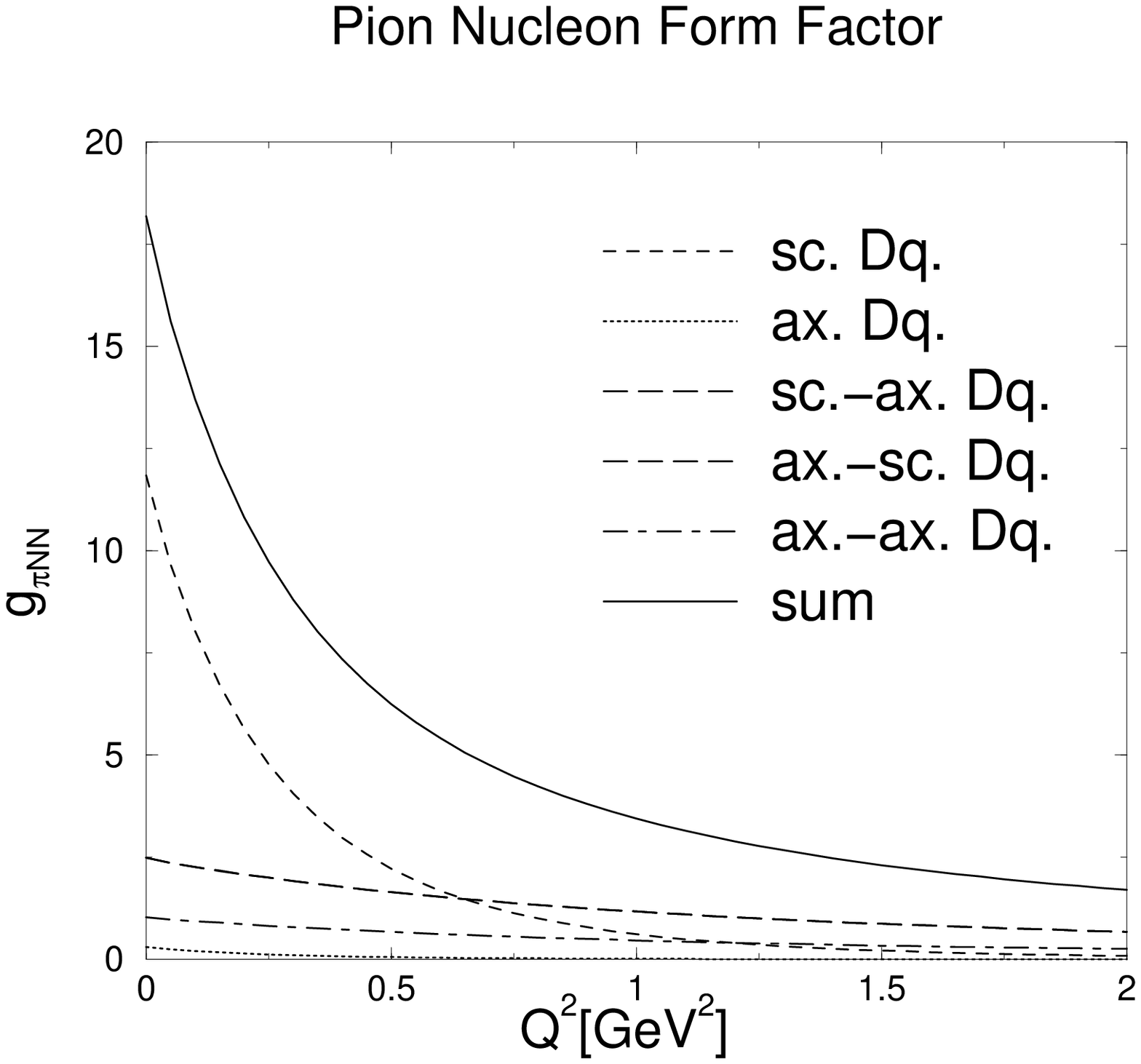,width=6.0cm,height=6.0cm}
\hspace{1cm}
\epsfig{file=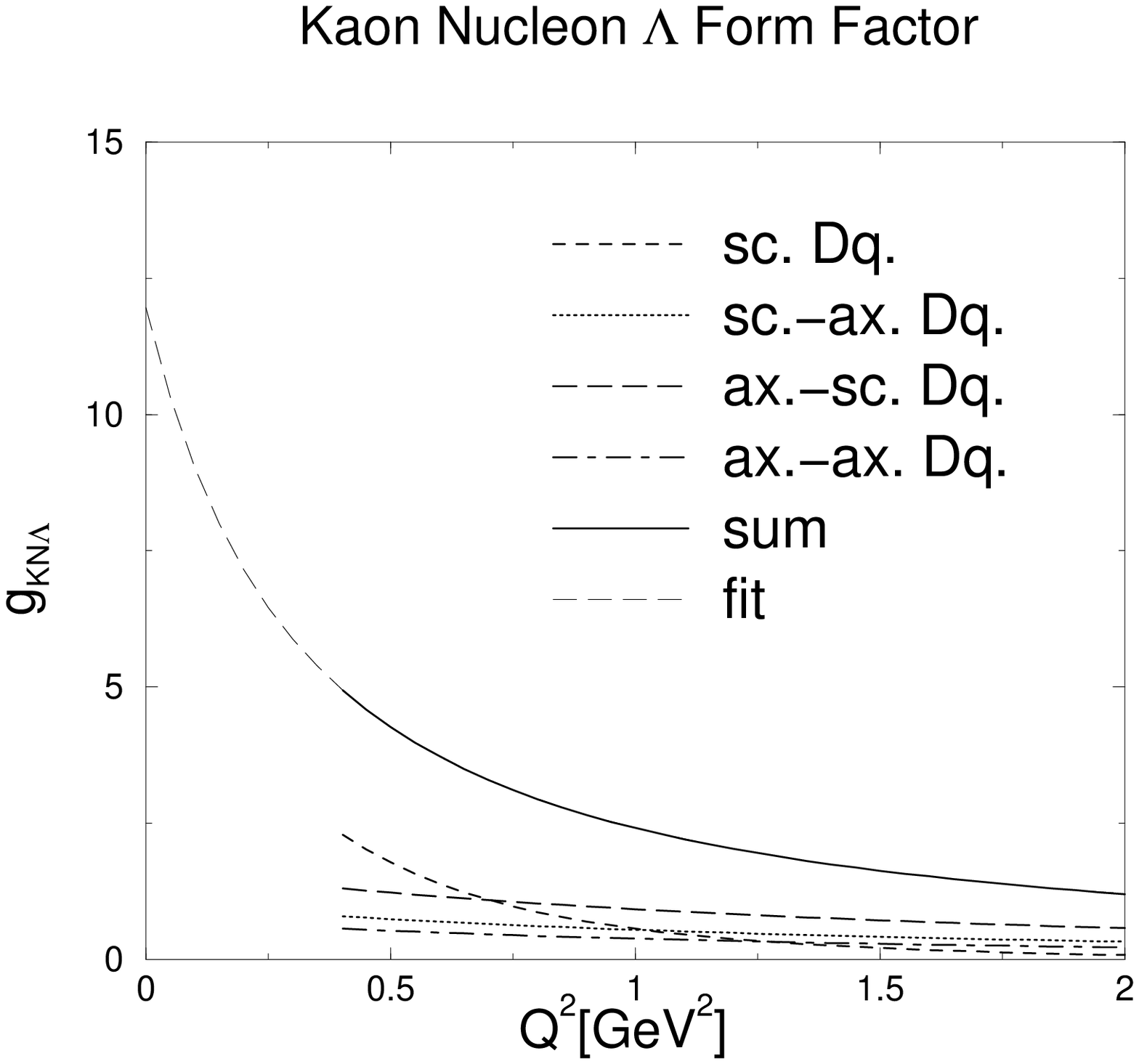,width=6.0cm,height=6.0cm}
}
\caption{\sf \label{g_cont}
The distinct contributions to the form factors $g_{\pi NN}$ and 
$g_{KN\Lambda}$ calculated using parameter set V.}
\end{figure}

For both form factors we observe a qualitative difference between 
calculations with or without axialvector diquarks included. Whereas 
for all parameter sets with only scalar diquarks the computed 
pion--nucleon form factor very well reproduces the experimental data, 
we find that for those sets that include axialvector diquarks the computed 
form factor overestimates the data. 
This could be  
due to the omission of subdominant amplitudes in the meson--quark 
vertex. We performed a rough estimate of the influence of the first
subleading amplitude by using a simple parametrization and indeed found negative 
corrections to the pion--nucleon form factor of about 30\%.
Future calculations should include these contributions in a selfconsistent way.
On the other hand in each of the subsets (I--IV) and (V--VI), the absolute
value of the couplings at $Q^2=0$ and the respective slope are almost
independent of the parameter sets and even of the propagator type.
To further analyze the structure of the form factors we have 
disentangled the various contributions in figure~\ref{g_cont}. This 
figure shows that at small positive $Q^2$ the contribution from the 
coupling of the meson to the scalar quark is clearly dominating, whereas 
for larger momenta the diquark contributions take over. 

As the form factors serve as input for later calculations we have 
conveniently fitted our numerical results to rational functions allowing, however, for 
a non--integer exponent,
\begin{equation}
g_{\phi B B^\prime}(Q^2)=
g_{\phi B B^\prime}\left(\Lambda_{\rm BF}^2\right)
\left(\frac{\Lambda^2+\Lambda_{\rm BF}^2}
{\Lambda^2+Q^2}\right)^\rho\, .
\label{fitsff}
\end{equation}
The result of this procedure are summarized in table~\ref{form_tab}. To
compare with results of other model calculations we have additionally 
fitted our results to monopole form factors. This has yielded scales $\Lambda$ 
between $200-300{\rm MeV}$ for the parameter sets with only scalar 
diquarks and scales around $500{\rm MeV}$ for the sets including 
axialvector diquarks. As can be seen from figure \ref{g_cont} the additional
contributions fall substantially slower than the scalar diquark one 
and become dominant for large $Q^2$. This effect can be interpreted as 
`hardening' of the form factor. In agreement with the results 
from ref.~\cite{Bloch:2000rm} our form factors are much softer than 
those usually substituted in one--boson--exchange potential models for 
production processes. Those empirical scales for the monopole form 
are larger than 
$1300{\rm MeV}$~\cite{Machleidt:1987hj}. However, other theoretical 
approaches, {\it e.g.} lattice measurements or QCD--sum rule calculations 
indicate a monopole behavior with much smaller scales, {\it cf}. 
ref.~\cite{Meissner:1995ra} and references therein. 

Our prediction for $g_{KN\Lambda}(Q^2=0)$ is comparable to those found 
of QCD-sum rule or Skyrme model calculations but somewhat smaller than 
the chiral bag model result, {\it cf}. ref.~\cite{Jeong:1999jh} and 
references therein. Extrapolating our $g_{KN\Lambda}(Q^2)$ to the
kaon mass shell $Q^2=-M_K^2$ yields values in the range 
$16.3 \le g_{KN\Lambda} \le 19.3$. This is slightly above the ballpark 
of the numbers extracted from experiment~\cite{Timmermans:1995qa}.

\begin{figure}
\centerline{
\epsfig{file=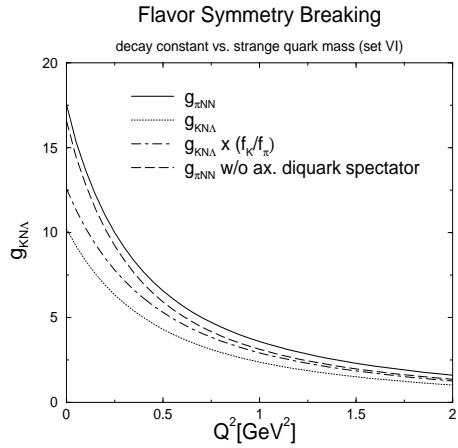,width=6.0cm,height=6.0cm}
}
\caption{\sf \label{g_flbreak}
Analysis of flavor symmetry breaking in $g_{KN\Lambda}$.}
\end{figure}

The comparison between $g_{\pi NN}$ and $g_{KN\Lambda}$
suggests three different scenarios of $SU(3)$-flavor symmetry breaking
that are illustrated in figure~\ref{g_flbreak}. The most obvious 
symmetry breaking stems from different quark mass parameters
$m_u\ne m_s$. Secondly, due to the flavor algebra the process 
in which the axialvector diquark acts as a spectator, only
contributes to $g_{\pi NN}$. Finally there are different decay
constants $f_K\ne f_\pi$ that factorize in the meson--quark vertices.
These three effects cause $g_{KN\Lambda}(Q^2)<g_{\pi NN}(Q^2)$
independently from the type of the propagator or adopted model 
parameters. We see from figure~\ref{g_flbreak} that at moderate and
large $Q^2$ the different decay constants dominate the symmetry 
breaking effects. For small $Q^2$ the mass differences are essential.

\begin{table}[t]
\begin{tabular}[b] {rrrrrrrr}
 & I & II & III & IV & V & VI & expt. \\
 diquark:
 & \multicolumn{4}{c}{scalar} & \multicolumn{2}{c}{scalar and}\\
 & & & & & \multicolumn{2}{c}{axialvector} \\ \hline \\
 propagator   & 2 & 1 & 1 & 3 & 1 & 3 \\
 type ($f_i$)  \\ \hline \\
 $g_{\pi NN}$ & 14.0 & 13.4 & 14.3 & 14.0 & 18.2 & 17.5 & 13.4 \\
 exponent & 7.8 & 5.1 & 5.1 & 5.2 & 1.4 & 1.6 \\
 scale [MeV]  & 1327 & 1106 & 1132 & 1252 & 650 & 778 \\ \hline \\
 $g_{KN\Lambda}$ & 7.98 & 7.39 & 8.25 & 8.12 & 11.97 & 10.23  \\ 
 exponent & 10.6 & 5.9 & 5.8 & 6.4 & 1.3 & 2.0  \\
 scale [MeV] & 1786 & 1368 & 1391 & 1554 & 642 & 950  
\end{tabular}
\caption{\sf \label{form_tab}
Numerical results for the absolute values of the couplings
$g_{\pi NN}$ and $g_{KN\Lambda}$ at zero squared momentum and for the
best rational fits to the curves. The parameter sets are defined 
in table~\protect\ref{para_tab}. The entries `exponent' and `scale'
refer to the variables $\rho$ and $\Lambda$ in the fit~(\protect\ref{fitsff}).}
\end{table}

\subsection{Associated Strangeness Production}
\label{prod_proc_numerics}

The COSY--TOF collaboration has measured the cross section and the
polarization for $pp \rightarrow pK\Lambda$ at $55{\rm MeV}$ and 
$138{\rm MeV}$ above threshold~\cite{COSY-TOF}. In addition 
there are also data for the depolarization tensor $D_{NN}$ from 
the DISTO collaboration at SATURNE~II~at an
excess energy of $E=431{\rm MeV}$~\cite{Balestra:1999br}. This tensor 
is an especially interesting observable. Eventually $D_{NN}$ might 
provide further information on the spin structure of the nucleon because
it describes the transport of spin from the initial to the final 
states ({\it cf.} appendix~\ref{ap_ASPRO} for appropriate definitions). 

\begin{figure}
\centerline{
\epsfig{file=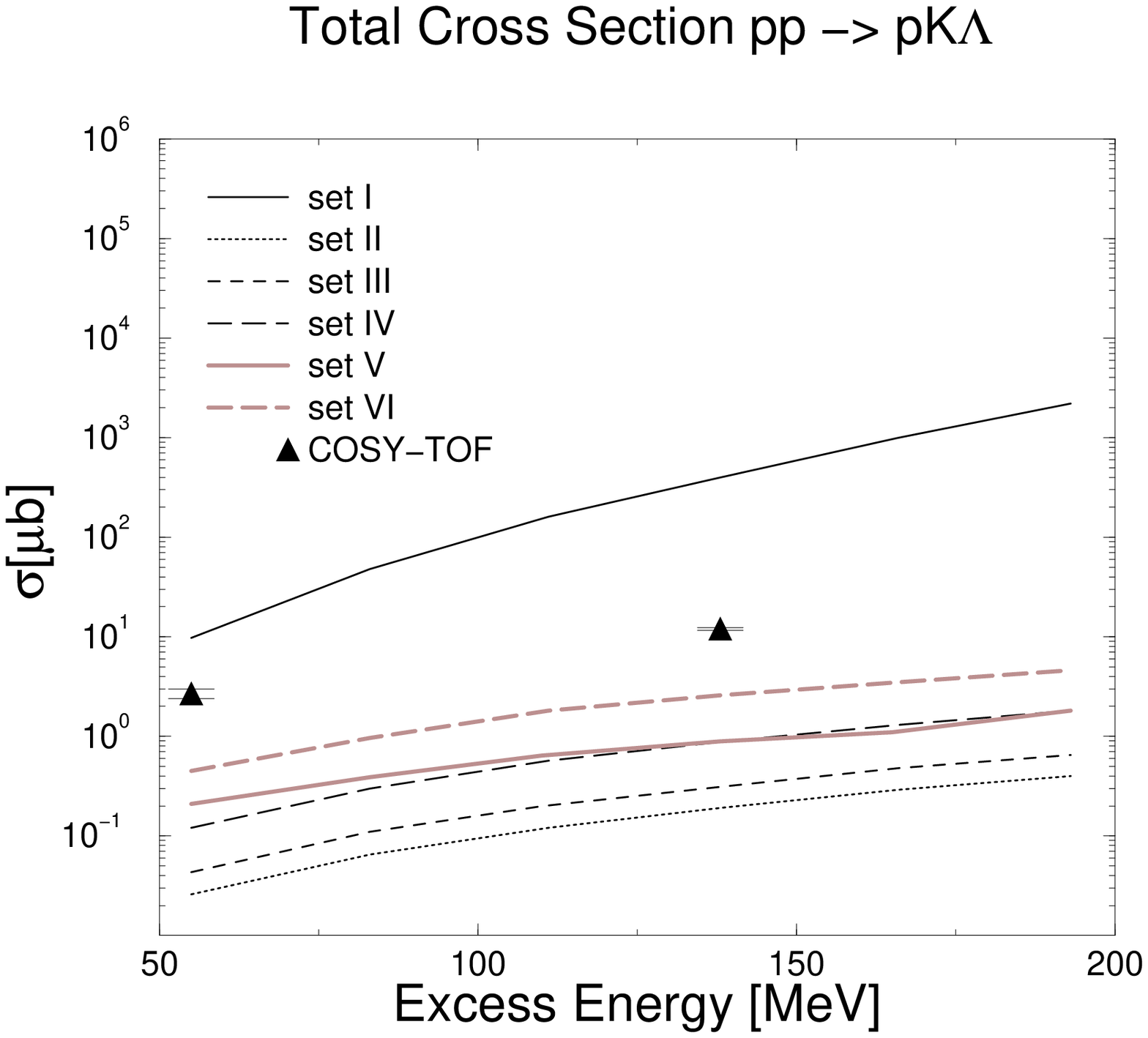,width=6.0cm,height=6.0cm}
\hspace{1cm}
\epsfig{file=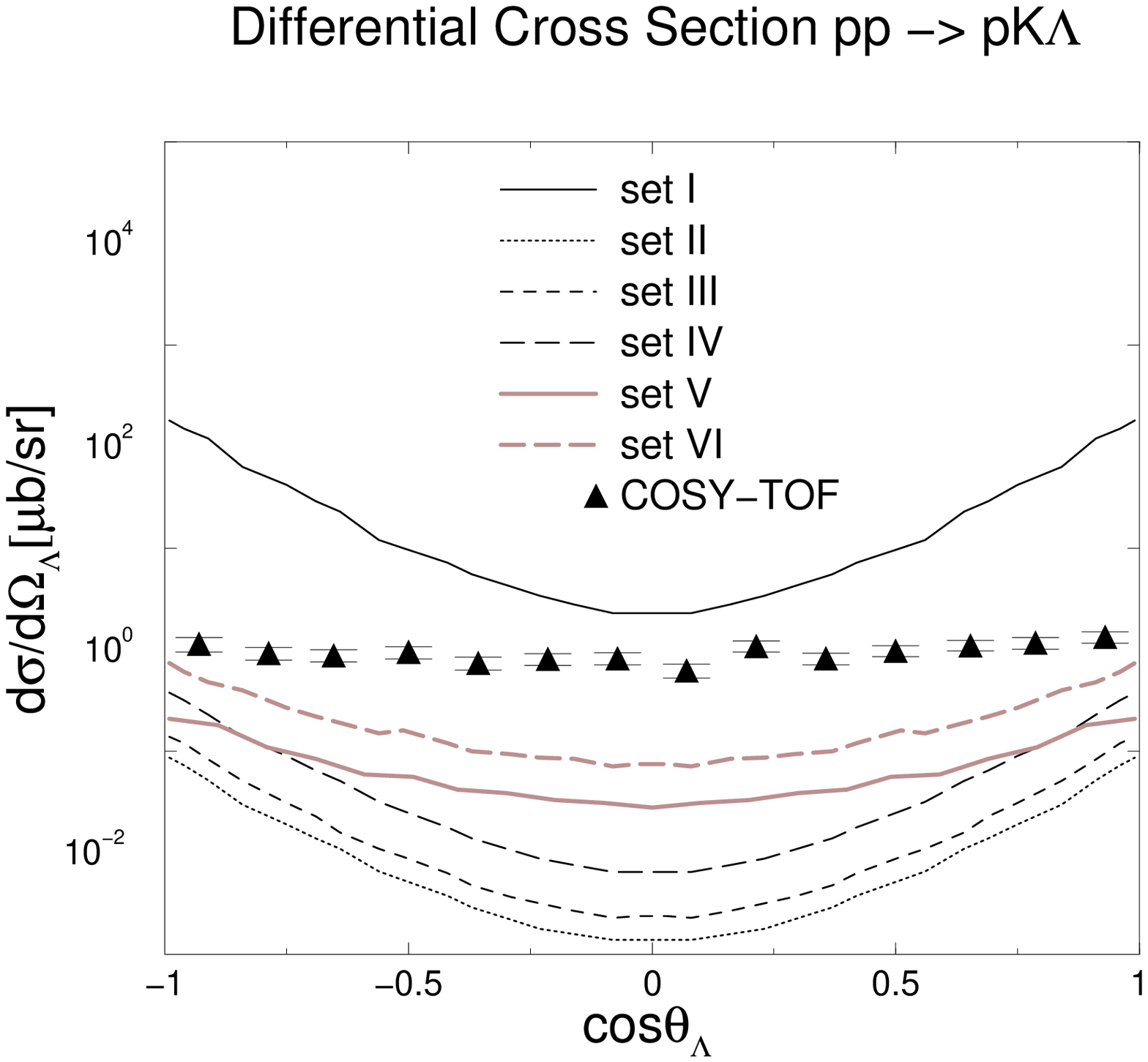,width=6.0cm,height=6.0cm}
}
\caption{\sf \label{pppkl_wirk}
Total and differential cross section of the process $pp \rightarrow
pK\Lambda$. The differential cross section (right panel) is considered
at an excess energy of $138{\rm MeV}$.}
\end{figure}

Our numerical results for associated strangeness production are shown in 
figure~\ref{pppkl_wirk}. The comparison with the empirical data 
clearly shows that the propagator with an exponential dressing 
function (set I) yields unacceptable results. As discussed in 
section~\ref{relevance} the mechanism is that by increasing the
beam momentum larger timelike momenta appear in the loop propagators
and hence the cross sections suffer an exponential enhancement. This 
effect is most strongly pronounced at forward and backward angles in 
the differential cross section.

\begin{figure}
\centerline{
\epsfig{file=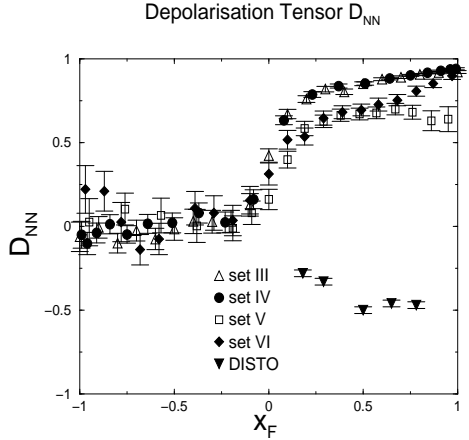,width=6.0cm,height=6.0cm}
}
\caption{\sf \label{pppkl_dnn}
The depolarization tensor $D_{NN}$ as a function 
of the Feynman variable $x_F$ which measures the ration of the actual
momentum of the $\Lambda$ projected onto beam direction divided by
the greatest possible one. The error bars on our numerical results
represent the statistical error of the Monte Carlo integration.}
\end{figure}

All propagators that do not involve the exponential dressing function
underestimate the cross section for $pp\to pK\Lambda$ considerably. Only 
the parameter set VI can be considered to be at the right order of magnitude.
Generally we find that the inclusion of axialvector diquarks improves the 
agreement with the data. This is not only the case for the total cross 
sections but also for the shape of the differential ones. The huge dip that
arises for the parameter sets I-IV at directions perpendicular to the 
beam axis is considerably damped by the axialvector diquark contributions 
although it is still too deep.

The two distinct contributions to the cross section that can be
characterized as being associated with pion or kaon exchange
({\it cf.} figure~\ref{pppkldiag}) lead to significant interference
cancellations for the depolarization tensor $D_{NN}$. The kaon exchange 
processes generate the outgoing $\Lambda$ in the form factor part
of the diagrams. Here the $\gamma_5$-structure of the kaon vertex leads to
a spin flip from the incoming proton to the $\Lambda$ because of
parity conservation. Therefore kaon exchange diagrams provide negative
contributions to the polarization tensor. In the pion exchange diagrams,
however, the outgoing $\Lambda$ is generated by the handbag part of the 
diagram. If both mesons were on--shell a spin flip of the quarks at each 
vertex would result in parallel spins of the incoming proton and the 
outgoing $\Lambda$. This would be a positive contribution to $D_{NN}$. 
Due to the off--shellness of the exchanged meson some small negative 
contributions arise. The actual magnitude depends on the particular 
kinematical situation considered. In essence, the depolarization tensor 
is controlled by the size of pion and kaon exchange contributions 
and in particular by the phases of the diagrams. 
These phases are completely controlled by the kinematics of the process.
We note that in other model calculations these phases are either
adjusted~\cite{Laget:1991jk} or interference terms were completely 
omitted~\cite{Sibirtsev:2000ut}.

For large negative values of the Feynman--parameter $x_F$ the outgoing 
$\Lambda$ is dominantly produced by the unpolarized target proton. 
This causes the depolarization tensor to vanish. According to 
figure~\ref{pppkl_dnn} our model calculations reproduce that feature.
For $x_F>0$ we obtain a sizable and positive depolarization tensor. This 
results from the fact that the leading contribution to the process 
stems form the diagram \#1 in figure~\ref{pppkldiag}. Here the pion 
couples to the quark while the diquark acts as spectator. As discussed
above such diagrams mainly produce $\Lambda$--spins that are parallel to 
the spin of the incoming proton. On the contrary the experimental
results suggest that the main contribution should stem 
from kaon exchange~\cite{Balestra:1999br}. We note, however, that this 
obvious discrepancy between theory and experiment has been found in 
other model calculations as well, {\it cf}. 
refs.~\cite{Sibirtsev:2000ut},\cite{Liang:1997rt}. 

\subsection{Kaon Photoproduction}
\label{prod_proc}

Here we will discuss our numerical results for the process
$\gamma p\to K\Lambda$. The technical details that enter this
calculation are given in appendix~\ref{app_product}.

In figure~\ref{tot_sig} we display the total cross section
$\sigma(\gamma p\to K\Lambda)$ as a function of the photon energy $E$. 
We observe that the parameter sets (II--IV) predict cross sections that 
are comparable with the experimental data. These model calculations do 
not include axialvector diquarks. Once these degrees of freedom are 
taken into account (sets V and VI), the cross--section is overestimated 
by about a factor four. For the five sets II--VI we find that the total
cross section is strongly dominated by the 
kaon--exchange diagram ({\it cf.} figure~\ref{kp_photo_diag})
while the handbag--type diagrams can almost be neglected.\footnote{It is
interesting to note that even for the handbag diagram alone results obtained
with a Ball-Chiu or bare photon quark vertex, respectively, differ by at most a
few percent.} As  only a single diagram contributes interference does not occur 
and it is obvious that the model calculations do not reproduce the 
dip in the energy region $1.1{\rm GeV}\le E \le 1.4{\rm GeV}$. Tuning the model
propagators such that the two diagrams are of equal importance this dip could be
reproduced \cite{Alk00}.
\begin{figure}[h]
\centerline{
\epsfig{file=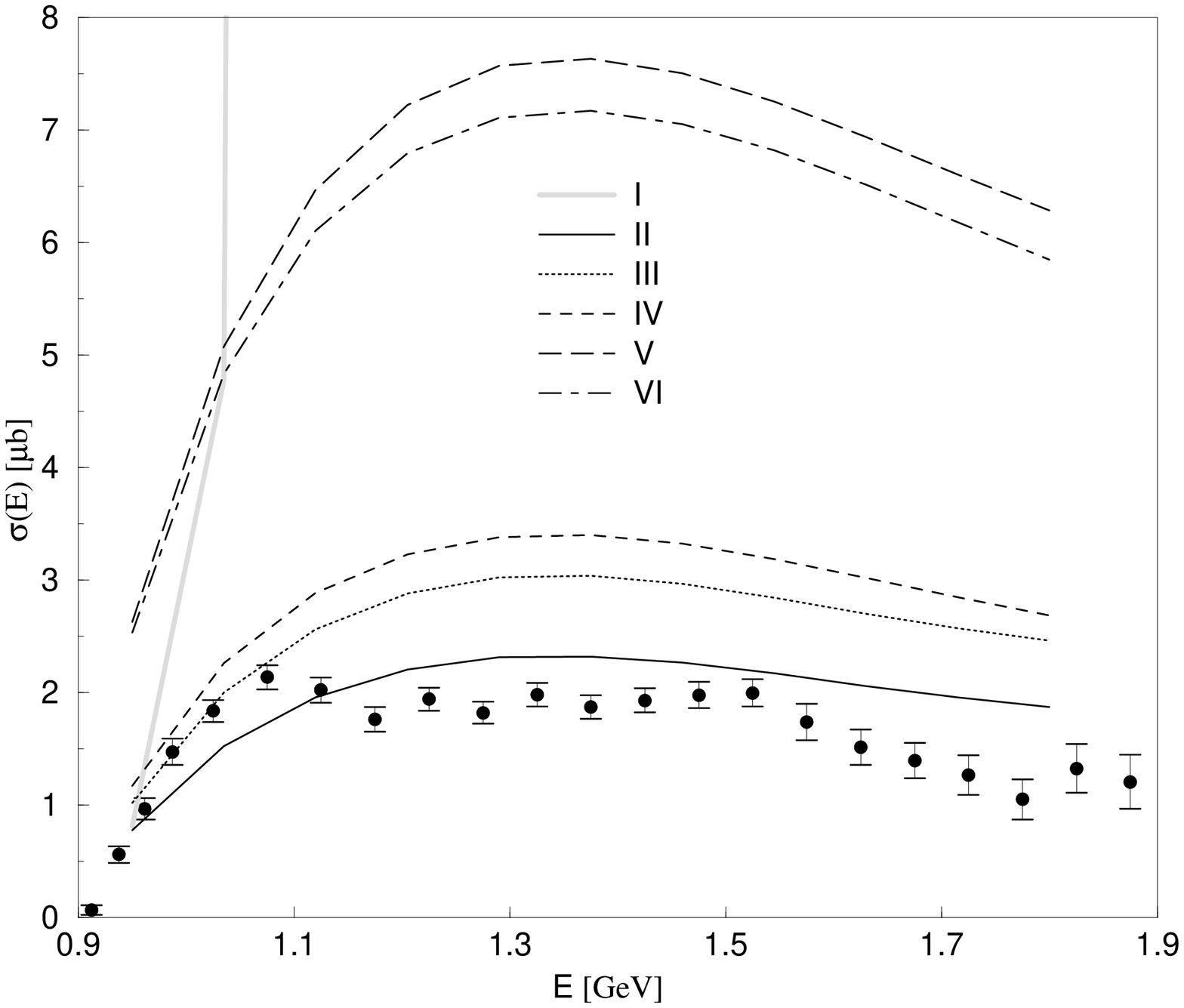,width=6.0cm,height=6.0cm,angle=270}
\hspace{1cm}
\epsfig{file=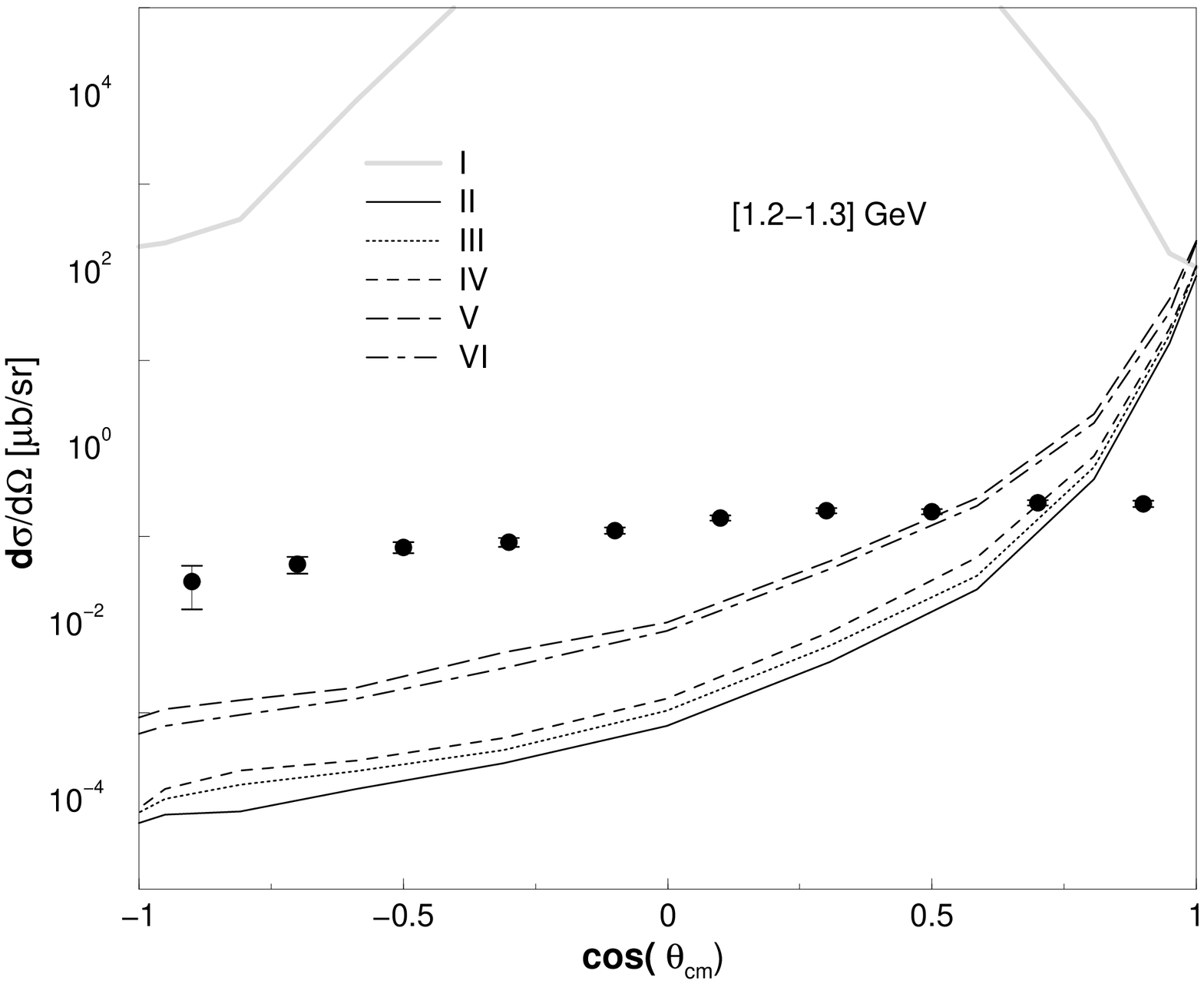,width=6.0cm,height=6.0cm,angle=270}
}
\caption{\sf \label{tot_sig}
The total cross section for kaon photoproduction as a function of the 
incident photon momentum $E$ (left panel) and the differential cross 
section for kaon photoproduction averaged in the energy bin 
$[1.2,1.3]{\rm GeV}$. The parameter sets (I-VI) are defined
in table~\protect\ref{para_tab}. The experimental data are 
taken from ref.\protect\cite{Tran:1998qw}.}
\end{figure}
As it has been the case for the associated strangeness production
we find that utilizing an exponential dressing function (set I) 
widely overestimates the experimental data. In this case actually the
handbag diagrams dominate while the kaon exchange contributions
are comparatively tiny. Figure~\ref{tot_sig} clearly shows that the large 
disagreement of the model results with the data certainly is not a 
fine--tuning problem. Rather we must conclude that the comparison 
with data rules out propagators that strongly rise in the timelike region
as the one dressed by an exponential function does.
Contrary to the case considered above
the kaon exchange diagram exceeds the handbag diagram by
almost one order of magnitude for the parameter sets II--IV.
Possible corrections from subleading meson amplitudes in the
kaon--quark vertex might decrease the strength of the
kaon--nucleon--$\Lambda$ form factor in a similar fashion as they do
for the pion--nucleon form factor. 
 
In figure~\ref{tot_sig} we also present the differential cross 
section in the energy interval $1.2{\rm GeV}\le E \le 1.3{\rm GeV}$
as a function of the angle between the momenta of the initial 
proton and the final kaon in the center of mass frame, {\it cf.}
appendix~\ref{app_product}. Although the model calculations reproduce 
the empirical increase of the differential cross section as 
${\rm cos}\,\theta_{cm}$ goes
from minus to plus unity, the increase appears to be overestimated.
For those parameter sets (II--VI) for which the resulting cross 
sections are dominated by the kaon exchange diagrams the predicted
differential cross sections turn out too small in the backward 
scattering region while they are too big in the opposite 
direction. As a result the total cross section agrees with the
empirical data reasonably well. Again, the exponential type 
propagators yield differential cross sections that are way off
the data and we repeatedly conclude that this type of propagators
is ruled out.

We have also computed the asymmetries that are defined in 
eqs~(\ref{p_asym})--(\ref{s_asym}). We find that they essentially
vanish for the model propagators that we consider reasonable, {\it i.e.}
sets II--VI. Although the model calculation correctly predicts
that the polarized photon asymmetry $\Sigma$, see eq~(\ref{s_asym}),
is positive for ${\rm cos}\,\theta_{\rm cm}<0$ and negative 
otherwise, the absolute values are off by several orders of
magnitude. Only when substituting propagators that are 
characterized by the exponential dressing function the 
predicted asymmetries roughly agree with the empirical
data. However, we have discarded already that propagator for other 
reasons given above.

\section{Conclusions}

In this paper we have considered baryons as fully relativistic bound states  
of quarks and separable quark--quark correlations, {\it i.e.}, diquarks. The
main purpose of this study has been to utilize empirical information in order
to restrict the structure of the propagators that model confined quarks and 
diquarks. These propagators enter the four--dimensional Bethe--Salpeter
equations from which we have computed the mass eigenvalues and wave functions
that are associated with physical baryons. 

The full covariance of the model wave functions allows us to unambiguously 
calculate form factors up to momentum transfers of several ${\rm GeV}$. For 
spacelike momenta the empirical form factors can be very well reproduced 
with tree--level quark and diquark propagators~\cite{Oettel:2000jj}. On the 
other hand the description of processes involving timelike momenta is 
obscured by the presence of quark thresholds in the tree--level
propagators. Of course, these thresholds are unphysical and reflect the 
absence of confinement. It is thus appropriate to modify these tree--level
propagators in order to implement the confinement phenomenon. In this 
framework we have considered three qualitatively different cases:
In the first case, the tree--level poles at timelike real $p^2$ have 
been traded for a pair of complex--conjugate poles. In this case the
imaginary parts (and therefore thresholds) cancel. In the second scenario, 
the pole on the timelike real $p^2$ axis has been screened at the expense 
of an essential singularity for infinite timelike momenta. In the third 
case, we have emphasized the issue that the propagators should equal the 
tree--level ones for all complex values of $p^2$ as $|p^2| \to \infty$. 
Together with the condition that no poles occur this property enforces a 
non--analytic form. We have then investigated the phenomenological 
implications of either of these forms rather than attempting
to precisely reproduce the experimental data. Obviously,
those processes are most interesting whose computation involves timelike 
momenta of the order of one GeV entering the model propagators.
In diquark models the flavor algebra alleviates the calculation of 
processes with a $\Lambda$ hyperon in the final channel. We have therefore
focused on kaon photoproduction and associated strangeness production with
the photoproduction being, at least in principle, more sensitive to 
timelike momentum transfers.

The model parameters have been fixed by fitting the baryon
spectrum and the nucleon electromagnetic form factors.
Both, the results for the magnetic moments of the nucleon and those
for the ratio $G_E/G_M$ show that it is important to include contributions
from the axialvector diquark. The strong form factors
$g_{\pi NN} (Q^2)$ and $g_{KN\Lambda}(Q^2)$ for spacelike momenta $Q^2>0$
depend on the amount of admixture of axialvector diquarks in the baryon
wave functions.
Our numerical result for
$g_{\pi NN} (0)$ using both,
scalar and axialvector diquarks, overestimates the empirically determined
value by approximately~30\%.
A possible reason for this discrepancy is the omission of
subleading amplitudes in the pion--quark vertex. Future calculations
should therefore include these contributions and the ones from
the kaon--quark vertex as well.
In any event, all these observables are almost insensitive to the
specific structure of the propagators. Therefore they do not provide
an adequate tool to distinguish between different parameterizations
of the confinement phenomenon.

The production processes, on the
other hand, strongly depend
on the form of the propagators in the
timelike region. In particular we have observed that the class of
propagators that is characterized by an exponential growth for large
timelike momenta overestimates the cross sections by orders of magnitude.
We have associated this failure to the dominance of the handbag--type
diagram. Apparently any quark propagator that for timelike momenta is
significantly more enhanced than the tree--level one immediately implies
the dominance of this diagram. The obvious conclusion is that those
propagators should be discarded. The other two forms of the propagators
have the potential to describe the data reasonably well. As mentioned,
we have omitted the so far undetermined subleading contributions
in the kaon exchange diagram that dominates the kaon photoproduction
amplitude. From the discussion of the pion form factor we have
sufficient reason to believe that the inclusion of such contributions will
favor parameter sets that contain axialvector diquarks.
Although the non--analytic form for the propagator could not be ruled out
by quantitative arguments we nevertheless think it should be discarded,
because it poses several fundamental problems related
with gauge and translational
invariance which we have detailed in the text. 
Therefore propagators that are characterized by pairs of
complex conjugated poles seem to be best suited for further studies.

\section*{Acknowledgments} 

\noindent
We thank C.\ D.\ Roberts, S.\ M.\ Schmidt and L.\ von Smekal 
for helpful discussions.

\noindent
This work has been supported by DFG (contracts Al 279/3-3 and
We 1254/3-1;4-2)
and  COSY (contract no.\ 41376610).

\begin{appendix}

\section{From the relativistic three--quark problem to the
diquark--quark model}
\label{3qreduce}

The six--point function $G(x_i,y_i)=\langle0|
T\prod_{i=1}^3q(x_i)\bar{q}(y_i)|0\rangle$ represents the
starting point for our study of the relativistic three--quark problem.
Here the variables $x_i$ and $y_i$ not only represent the space--time
coordinates but also include the discrete labels color, spin, and
flavor. The six--point function obeys the Dyson equation
\begin{equation}
G=G_0+G_0\otimes K \otimes G\, .
\label{G_Dyson}
\end{equation}
The entries of the Dyson equation~(\ref{G_Dyson}) are the disconnected
six--point function $G_0$ that describes the free propagation of
three quarks and the three--quark scattering kernel $K$ that contains
all two-- and three--particle irreducible diagrams. The symbol
``$\otimes$'' in eq~(\ref{G_Dyson}) denotes summation/integration over
all independent internal coordinates and labels. Unless explicitly stated
otherwise we will henceforth work in momentum space with Euclidean metric.
It is thus not necessary to introduce different symbols for momentum and
coordinate space objects.

A three--particle bound state with mass $M$ manifests itself as a pole
in the six--point function at $-P^2=M^2$ where $P=p_1+p_2+p_3$ is
the total four--momentum of the three--quark system. We may thus parameterize
the six--point function in the vicinity of the pole as
\begin{equation}
 G(k_i,p_i) \sim \frac{\psi(k_1,k_2,k_3)
\;\bar{\psi}(p_1,p_2,p_3)}{P^2+M^2} \; ,
 \label{bspole}
\end{equation}
where $\psi$ denotes the bound state wave--function. Substituting
this parameterization into the Dyson equation (\ref{G_Dyson})
and identifying residua, we find the homogeneous bound state equation
\begin{equation}
 \psi= G_0 \otimes K \otimes \psi  \quad
\Longleftrightarrow \quad
G^{-1}\otimes \psi=0\; .
 \label{3bpsi}
\end{equation}
Despite its simple appearance this equation is infeasible as neither
all two-- and three--particle graphs, $K$, nor the fully dressed
quark propagator, that is contained in $G_0$, are known. We will have to
resort to approximations that render the problem tractable and {\it a
posteriori} validate these approximations from the resulting bound state
properties.

\begin{figure}
 \begin{center}
   \epsfig{file=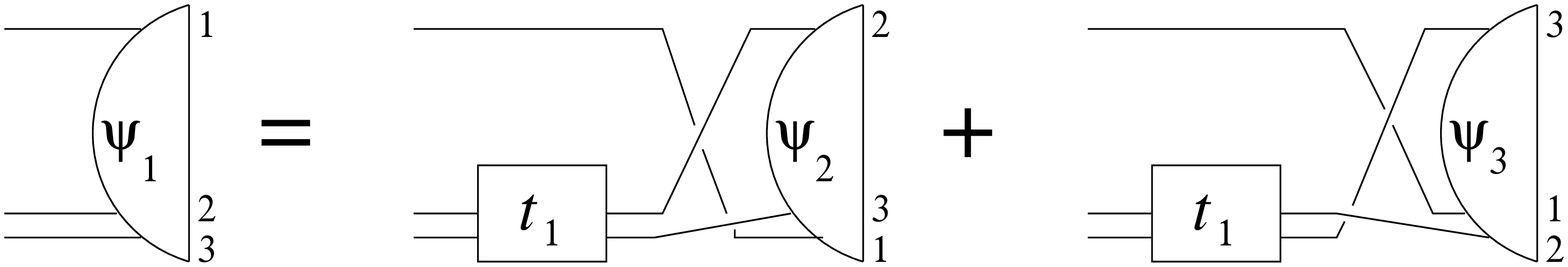,width=10cm,height=2cm}
 \end{center}
 \caption{\sf The Faddeev bound state equation for the component
          $\psi_1$. The  equations for $\psi_2$ and  $\psi_3$
          follow by cyclic permutation of the particle indices.}
 \label{faddeev_fig}
\end{figure}

\begin{figure}[b]
 \begin{center}
   \epsfig{file=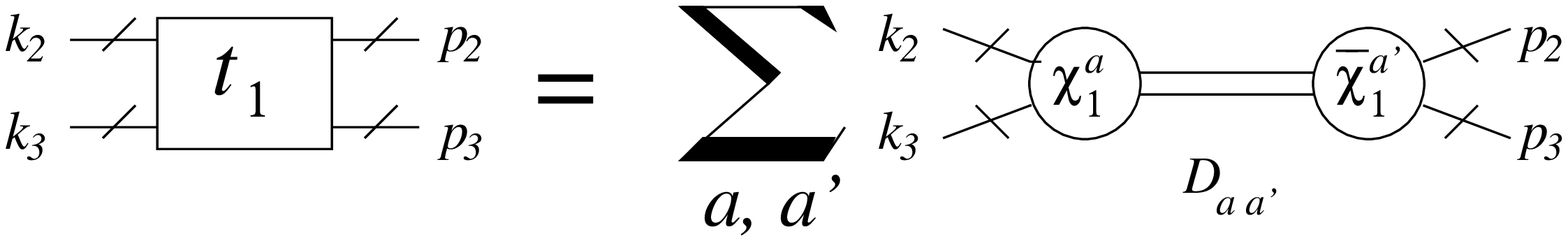,width=8cm,height=1.5cm}
 \end{center}
 \caption{\sf The separable matrix $t_1$. Also indicated is
the amputation of the external legs in the quark propagators.}
 \label{tsep_fig}
\end{figure}

The problem greatly simplifies when discarding all three--particle
irreducible graphs from the interaction kernel $K$. The kernel may then
be written as the sum of three two--quark interaction kernels,
\begin{equation}
K\,=\,K_1+K_2+K_3 \; .
\label{k2pi}
\end{equation}
We adopt the notation that the subscript of $K_i$ refers to the spectator
quark $q_i$. The respective interacting quark pair is $(q_j,q_k)$ with
the three labels ($i,j,k$) being a cyclic permutation of $(1,2,3)$.
These two--quark interaction kernels govern the Dyson equation for the
two--quark correlation functions, $g_i$:
\begin{equation}
g_i= G_0 +G_0 \otimes K_i \otimes g_i \; .
\label{g_i}
\end{equation}
As the appearance of the free six--point function $G_0$ suggests we
have defined both $g_i$  and $K_i$ in the three--quark space. This is
easily accomplished by attaching the propagator\footnote{In the framework
of these integral equations we factorize the momentum conservation
$\langle p_i|S_i|p_i^\prime\rangle=(2\pi)^4S(p_i)
\delta^4(p_i^\prime-p_i)$. Here $S(p)$ is the ordinary Dirac propagator
while $S_i$ refers to an operator in functional space. We adopt
analogous conventions for the other operators.} $S_i$ of the spectator
quark to $g_i$ and
its inverse $S_i^{-1}$ to $K_i$. Expressing eq~(\ref{g_i}) as
$G_0 \otimes K_i=1-G_0\otimes g_i^{-1}$ allows us to replace any of
the three operators $G_0 \otimes K_i$ in the bound state
equation~(\ref{3bpsi}),
\begin{equation}
 \psi = \left(1-G_0\otimes g_i^{-1}\right)\otimes\psi
        +G_0\otimes(K_j+K_k)\otimes  \psi \quad
\Longleftrightarrow \quad
 \psi =g_i \otimes (K_j+K_k)\otimes  \psi \; .
\end{equation}
To further elaborate this form of the bound state equation we
define the matrix $\hat{t}_i$ via
\begin{equation}
 g_i= G_0 + G_0 \otimes \hat{t_i} \otimes G_0 \; .
 \label{t1_def}
\end{equation}
This reflects the amputation of the external quark legs from $g_i$ after
having separated the non--interacting contribution. As already
mentioned we carry along factors of the quark propagators and its
inverse to formulate the problem in the three--quark space. For later
convenience we therefore define
\begin{equation}
t_i=\hat{t_i}\circ S_i
\label{t2_def}
\end{equation}
with the additional factor removed. We have introduced the symbol
``$\circ$'' to denote simple multiplications without any contractions
because the so--combined operators act on different quarks.
Finally we introduce the
Faddeev components $\psi_i$ by
\begin{equation}
 \psi_i = G_0\otimes K_i \otimes \psi \; .
\end{equation}
Upon rewriting the definition~(\ref{t1_def}) for $\hat{t_i}$ as
$g_i\otimes G_0^{-1}=1+G_0\otimes\hat{t}_i$  we find the
bound state equation
$\psi=\psi_j+\psi_k+G_0\otimes\hat{t}_i\otimes(\psi_j+\psi_k)$
and thus
\begin{equation}
 \psi_i = G_0 \otimes \hat t_i \otimes (\psi_j+\psi_k)
   = (S_j\circ S_k) \otimes t_i\otimes (\psi_j+\psi_k) \;.
 \label{Faddeev}
\end{equation}
These are the famous Faddeev bound state equations relating the Faddeev
component $\psi_i$ to $\psi_j$ and $\psi_k$. The graphical representation
of these equations is shown in figure~\ref{faddeev_fig}. These equations
embody the full two--quark correlation function $t_i$ instead of the
kernel $K$.
The relativistic Faddeev equations are a set of coupled four--dimensional
integral equations and represent a considerable simplification over the
original eight--dimensional integral equation problem defined in
eq~(\ref{3bpsi}).
Unfortunately the Faddeev components $\psi_i$ still depend on the two
relative momenta between the three quarks. Expanding these components in
Dirac space~\cite{Carimalo:1993ia} yields an intractable number of
coupled integral equations. We therefore further simplify the bound state
problem. Denoting the incoming and outgoing momenta by respectively
$k_i$ and $p_j$ we assume that the two--quark correlations $t_i$
do \underline{not} depend on any of the scalar products
$k_i\cdot p_j$ that connect momenta of the incoming and outgoing
channels. This assumption allows us to expand $t_i$ in terms
of separable correlations in the two--quark subspace that is
characterized by $j,k\ne i$ and $j\ne k$,
\begin{equation}
t_i(k_j,k_k;p_j,p_k)=\sum_{a,a^\prime} \chi_i^a(k_j,k_k)\;
D_{a,a^\prime}(k_j+k_k)\; \bar\chi_i^{a^\prime}(p_j,p_k)\; .
\label{sep_ass}
\end{equation}
We call these separable correlations ``diquarks'' and comprise the
various types together with their discrete quantum numbers within
the label~$a$. Note that the propagator $D_{a,a^\prime}(k_j+k_k)$ is
diagonal in the discrete quantum numbers $a$ and $a^\prime$ except for
the Lorentz indices for the axialvector diquark. Furthermore $\chi^a$
represents the vertex function of two quarks with a diquark.
Correspondingly $\bar \chi^a$ denotes the conjugate vertex function.
The expansion~(\ref{sep_ass}) is pictured in figure~\ref{tsep_fig}.

In a full solution to the Faddeev problem the $t_i$ will have to be
determined from the Dyson equation for $\hat{t}_i$
\begin{equation}
 \hat{t}_i = K_i + K_i \otimes G_0 \otimes \hat{t}_i \; ,
 \label{hatt_i}
\end{equation}
that follows from eq~(\ref{g_i}) and involves the kernel components $K_i$.
Rather than determining these vertices and the diquark propagators
from that Dyson equation we will adopt phenomenologically motivated
parameterizations for these quantities.

\begin{figure}[t]
 \begin{center}
   \epsfig{file=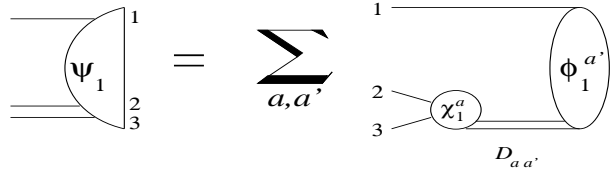,width=8cm,height=2.2cm}
 \end{center}
 \caption{\sf The {\em ansatz} (\protect{\ref{ansatz1}}) for the Faddeev
 component $\psi_1$ of the bound state wave function $\psi$ using effective
    baryon-quark-diquark vertex functions $\phi^a_1$.}
 \label{psisep_fig}
\end{figure}

Upon the separability assumption for the two--quark correlations
we continue to formulate a relativistic description of baryons
based on the Faddeev equations~(\ref{Faddeev}). In this approach
it is advantageous to introduce an effective vertex function,
$\phi^a_i$, for the interaction of the baryon with the quark
and the diquark. This vertex function depends only on the momentum of
the spectator quark, $p_i$, and the momentum, $p_j+p_k$ of the diquark
quasiparticle. Eventually this can be reexpressed as a dependence
on the relative momentum between quark and diquark, $\bar{p}_i$
as well as the total momentum $P$: $\phi^a_i=\phi^a_i(\bar{p}_i,P)$.
These dependencies are further detailed in
refs.~\cite{Thomas:1977,Ishii:1995bu}. In contrast to the
non--relativistic formulation we have some freedom in the
definition of the relative momentum. We may write
\begin{equation}
\bar{p}_i=p_i-\eta P=(1-\eta)p_i-\eta(p_j+p_k)\, ,
\label{relmom}
\end{equation}
where the parameter $\eta$ parameterizes the partition of this momentum
among the constituents. Of course, physical observables like the mass
of the bound state or form factors do not depend on this parameter
(up to numerical uncertainties).
The superscript in $\phi^a_i$ selects a diquark content, $a$, that builds a
baryon together with the quark of species $q_i$. A suitable {\it ansatz}
for the Faddeev components $\psi_i$ then reads
\begin{eqnarray}
 \psi_{i}^{\alpha\beta\gamma}&=&
S_i^{\alpha\alpha^\prime}S_j^{\beta\beta^\prime}
S_k^{\gamma\gamma^\prime} \sum_{aa^\prime}
\chi^a_{i,\beta^\prime\gamma^\prime}\;
D_{aa^\prime}\phi^{a^\prime}_{i,\alpha^\prime}\; ,
 \label{ansatz1}
\end{eqnarray}
where we have made the quark labels explicit. As usual, we sum over
doubly appearing indices. The quark label $i$ fixes the diquark
indices ($jk$). The momentum routing follows these indices as well as
the diquark labels $a$ and $a^\prime$. For further guidance we have
visualized this~{\it ansatz} in figure~\ref{psisep_fig}.
Noting that $G_0=S_i\circ S_j\circ S_k$ we find the
coupled integral equations for the effective vertex functions
\begin{equation}
 \phi^a_{i,\alpha}= \sum_{bb^\prime}
\left[\bar{\chi}^a_{i,\beta\gamma}S_k^{\gamma\gamma^\prime}
\chi^b_{j,\gamma^\prime\alpha}\right]\,
\left[D_{bb^\prime}S_j^{\beta\beta^\prime}
\phi^{b^\prime}_{j,\beta^\prime}\right]
\quad+\quad\big(j\longleftrightarrow k \big)\; ,
 \label{BS1}
\end{equation}
when inserting the {\it ansatz}~(\ref{ansatz1}) together with the
diquark parameterization~(\ref{sep_ass}) into the Faddeev
equations~(\ref{Faddeev}). In deriving eq~(\ref{BS1}) we have
utilized that the quark--diquark vertex functions are antisymmetric
under the exchange of the quark labels,
$\chi^a_{i,\beta\gamma}=-\chi^a_{i,\gamma\beta}$.
This feature is a consequence of the Pauli exclusion principle. We
have arranged the terms in eq~(\ref{BS1}) such as to exhibit the
similarity with the structure of Bethe--Salpeter equations. The first
term in square brackets represents a six--point function for quarks
that is governed by the exchange of a single quark between a quark
and a diquark. By coupling to the vertex function via the propagators
for quarks and diquarks it serves as the interaction kernel that
generates the Bethe--Salpeter equation for a bound state of quarks
and diquarks. Thus the Bethe--Salpeter equation sums the ladder--type
quark exchange diagrams between quarks and diquarks. Using the above 
definitions for total and relative momentum one arrives now at
the Bethe--Salpeter equation (\ref{BS2}).

\section{Decomposition of the diquark--quark Bethe--Salpeter amplitude}
\label{decomposition}

Here we will make explicit the full structure of the vertex functions
$\phi^a_i$  for the case of the  nucleon--quark--diquark vertex. For
identical quarks the nucleon--quark--diquark vertex functions
$\phi^a_i$ do not depend on the quark label $i$. The vertex
functions consist of a spinor in the case of a scalar diquark ($a=5$) and
a vector--spinor in the case of an axialvector diquark
($a\equiv \mu=1\dots 4$). Using positive energy spinors $u(P)$ with $P$
being the nucleon momentum, we define matrix--valued vertex functions
$\Phi=\pmatrix{\Phi^5\cr \Phi^\mu}$ via
\begin{equation}
  \phi^a(p,P)=\Phi(p,P) \; u(P)\; .
  \label{vertnucdef}
\end{equation}
Upon attaching quark and diquark legs to $\Phi$ we obtain the
matrix--valued Bethe--Salpeter wave functions
$\Psi=\pmatrix{\Psi^5\cr\Psi^\mu}$,
\begin{eqnarray}
 \tilde D (p_d) &:=& \pmatrix{D(p_d) & 0 \cr 0 & D^{\mu\nu}(p_d)}
\label{tildeD} \; ,\\
 \Psi(p,P)&=& \left[S(p_q) \circ \tilde D (p_d)\right]
 \Phi(p,P) \, .
 \label{wavenucdef}
\end{eqnarray}

We demand that the nucleon Faddeev amplitude eq~(\ref{ansatz1}) has
positive parity and describes positive--energy states. The latter
condition enforces that wave-- and vertex functions are
eigenfunctions of the positive--energy projector
$\Lambda^+=\frac{1}{2}(1-\imag\Pslash/M)$, {\em i.e.},
\begin{equation}
 \Phi=\Phi\;\Lambda^+\qquad {\rm and} \qquad \Psi=\Psi\;\Lambda^+\; .
\end{equation}
Using these constraints, the most general structure of $\Phi$
contains two amplitudes (scalar functions) $S_1(p,P)$ and $S_2(p,P)$
coupling to the scalar correlations and six amplitudes
$A_1(p,P),\dots,A_6(p,P)$ for the axialvector
correlations within the nucleon. Explicitly,
\begin{eqnarray}
 \label{vex_N}
\Phi^5 (p,P) &=& \sum_{i=1}^2 S_i(p^2,p\cdot P)\,
{\mathcal S}_i(p,P)\; ,
\nonumber \\
\Phi^\mu(p,P) &=&
\sum_{i=1}^6 A_i(p^2,p\cdot P) \; \gamma_5 {\mathcal A}_i^\mu(p,P)\; .
 \label{vertexdeco1}
\end{eqnarray}
The Dirac components ${\mathcal S}_1,\cdots,{\mathcal A}_6$ that obey
the positive energy condition are listed in table~\ref{components1}.
Also, these components have positive parity. We remark that the
wave--function $\Psi$ can be analogously expanded because it must obey
the same constraints as $\Phi$ does. In the nucleon rest frame the
individual components of $\Psi$ are eigenfunctions of the three--quark
spin and orbital angular momentum operators, respectively, when the
Faddeev amplitude is expanded within the
basis~(\ref{vertexdeco1})~\cite{Oettel:1998bk}. Thus, the amplitude $S_1$
describes the strength of an $s$--wave in the scalar channel while
$A_1$ and $A_3$ represent $s$--waves in the axialvector channel. There
also is a small $d$--wave component in the nucleon parameterized by $A_5$.
All amplitudes with even labels are relativistic (lower) components
associated with the above described amplitudes that have an odd label,
and these flavor components are absent in a nonrelativistic description.

\begin{table}
 \begin{center}
 \begin{math}
  \begin{array}{crc} \hline \hline\\
    \phantom{++}{\mathcal S}_1\phantom{++} & \Lambda^+ & \\[2mm]
    {\mathcal S}_2 & -i \,\hat \Slash{p}_T\, \Lambda^+  \\[2mm] \hline
    \\[-4mm]
    {\mathcal A}_1^\mu & \hat P^\mu\,\Lambda^+ \\[2mm]
    {\mathcal A}_2^\mu & -i \,\hat P^\mu\,\hat \Slash{p}_T \,\Lambda^+\\[2mm]
    {\mathcal A}_3^\mu & \frac{1}{\sqrt{3}}\,\gamma^\mu_T\, \Lambda^+ \\[2mm]
    {\mathcal A}_4^\mu & \frac{i}{\sqrt{3}} \,\gamma^\mu_T\,
        \hat \Slash{p}_T \,\Lambda^+ \\[2mm]
    {\mathcal A}_5^\mu & \sqrt{\frac{3}{2}}\left(\hat p^\mu_T \,\hat
        \Slash{p}_T - \frac{1}{3}\, \gamma^\mu_T \right) \Lambda^+ \\[2mm]
    {\mathcal A}_6^\mu & i \sqrt{\frac{3}{2}} \left( \hat p^\mu_T -
        \frac{1}{3} \,\gamma^\mu_T \,\hat \Slash{p}_T \right) \Lambda^+ \\
    \\ \hline \hline
  \end{array}
 \end{math}
 \end{center}
 \caption{\sf Basic Dirac components of the nucleon vertex function.
   The hat denotes normalized four--vectors, $\hat p=p/|p|$.
   For the complex on-shell nucleon momentum, we define
   $\hat P=P/\imag M$. The subscript ``$\,T\,$'' denotes the transversal
   component of a vector with respect to the nucleon momentum $P$,
   {\em e.g.} $p_T=p-(p\cdot \hat P) \hat P$.}
 \label{components1}
\end{table}

\section{Technical details for calculating production processes}
\label{app_product}

\subsection{Kaon Photoproduction $p\gamma \rightarrow \Lambda K$}
\label{app_KaPho}

\begin{figure}
  \begin{center}
     \epsfig{file=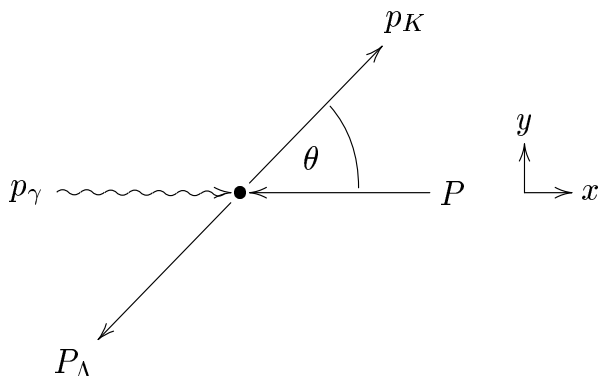,width=8cm,height=5cm}
     \caption{\sf Kinematics for kaon photoproduction 
       $p\gamma \rightarrow \Lambda K$ in the center of momentum frame. 
       The incoming proton and photon carry the momenta $P$ and 
       $p_{\gamma}$ respectively. The outgoing kaon and $\Lambda$ are 
       labeled by the momenta $p_K$ and $P_\Lambda$.}
     \label{kapho_kinme}
  \end{center}
\end{figure}

In this appendix we detail the calculation of the diagrams shown in 
figure~\ref{kp_photo_diag}. They provide the main contributions to 
the photoproduction process $p\gamma \rightarrow \Lambda K$.

We have performed the calculations in both the rest frame of the proton 
and in the center of momentum system (CMS). For the following discussion
we choose the CMS for definiteness. The momenta are defined 
according to figure~\ref{kapho_kinme}, that is
\begin{eqnarray}
P&=&(-E,0,0,\imag E_P)\label{kapho.extmom.1}
\, ,\quad
p_{\gamma}=(E,0,0,\imag E)
\nonumber \\
p_K&=&(|\vec{p}_K|\cos{\theta},|\vec{p}_K|\sin{\theta},0,\imag E_K)
\, ,\quad
P_\Lambda=(-|\vec{p}_K|\cos{\theta},
-|\vec{p}_K|\sin{\theta},0,\imag E_\Lambda) 
\label{kapho.extmom.4}
\end{eqnarray}
with
\begin{eqnarray}
E_P&=&\sqrt{M_P^2 + E^2}
\, ,\quad
E_K=\frac{1}{ 2(E+E_P) }
\left((E+E_P)^2 - M_{\Lambda}^2 + M_K^2 \right) 
\nonumber\\
|\vec{p}_K|&=& \sqrt{(E_K)^2-M_K^2} 
\, ,\quad
E_\Lambda=\sqrt{\vec{p}_K\hspace{0.5pt}^2+M_{\Lambda}^2}  \, .
\end{eqnarray}
The on--shell conditions and momentum conservation leave only two kinematical
variables undetermined. These are usually chosen to be the energy $E$ 
of the incoming photon and the angle~$\theta$ between the spatial momenta 
of the photon and the outgoing kaon.

The three diagrams in figure~\ref{kp_photo_diag} show the 
contributions to the transition amplitudes that we will discuss here.
The uncrossed `handbag diagram' translates into
\begin{equation}
A_1=\imag\int \frac{d^4 p}{(2\pi)^4}
\left\{\bar{\Phi}_\Lambda(p_f,P_\Lambda)S(p_+)
\Gamma_K (q,p_+)S(q)\right\}
\left\{\Gamma_{\gamma}(p_-,q)S(p_-)\Phi(p,P)D(p_d)\right\}
\end{equation}
with the momentum routing described in eq~(\ref{A1.mom.1}), see also 
figure~\ref{kp_photo_diag}. The crossed `handbag diagram' corresponds 
to the expression
\begin{equation}
A_2=\imag\int \frac{d^4 p}{(2\pi)^4}
\left\{\bar{\Phi}_\Lambda(p_f,P_\Lambda,)S(p_+)
\Gamma_{\gamma}(q,p_+)S(q)\right\} 
\left\{\Gamma_K(p_-,q)S(p_-)\Phi(p,P)D(p_d)\right\}\, .
\label{A2.mom.1}
\end{equation}
The definitions for $q$ and $p_+$ have changed as compared to 
the momentum routing for the amplitude $A_1$ given in 
eq~(\ref{A1.mom.1}). In eq~(\ref{A2.mom.1}) we have instead:
\begin{equation}
  q   = p_- - p_K   \quad{\rm and}\quad
  p_+ = q   + p_{\gamma}\, ,
\end{equation}
with all other momentum definitions unchanged. The amplitude 
corresponding to the tree level diagram arising from kaon exchange
is given by
\begin{equation}
  A_3 = \left( \bar{u}_{\Lambda} [ g_{KN\Lambda} \gamma_5 ] u_p \right)
        \frac{1}{q^2 + M_K^2}
        \left( \Gamma^{\nu} \epsilon_{\nu} \right)\; .
\end{equation}
Here $g_{KN\Lambda}$ represents the strong form factor that has been 
discussed in subsection~\ref{strongFF} and $M_K$ is the kaon mass. The 
photon polarization is denoted by $\epsilon_{\nu}$ while $\Gamma^{\nu}$ 
refers to kaon--photon vertex containing the electromagnetic kaon form factor
\begin{equation}
\begin{minipage}{2cm}
\epsfig{file=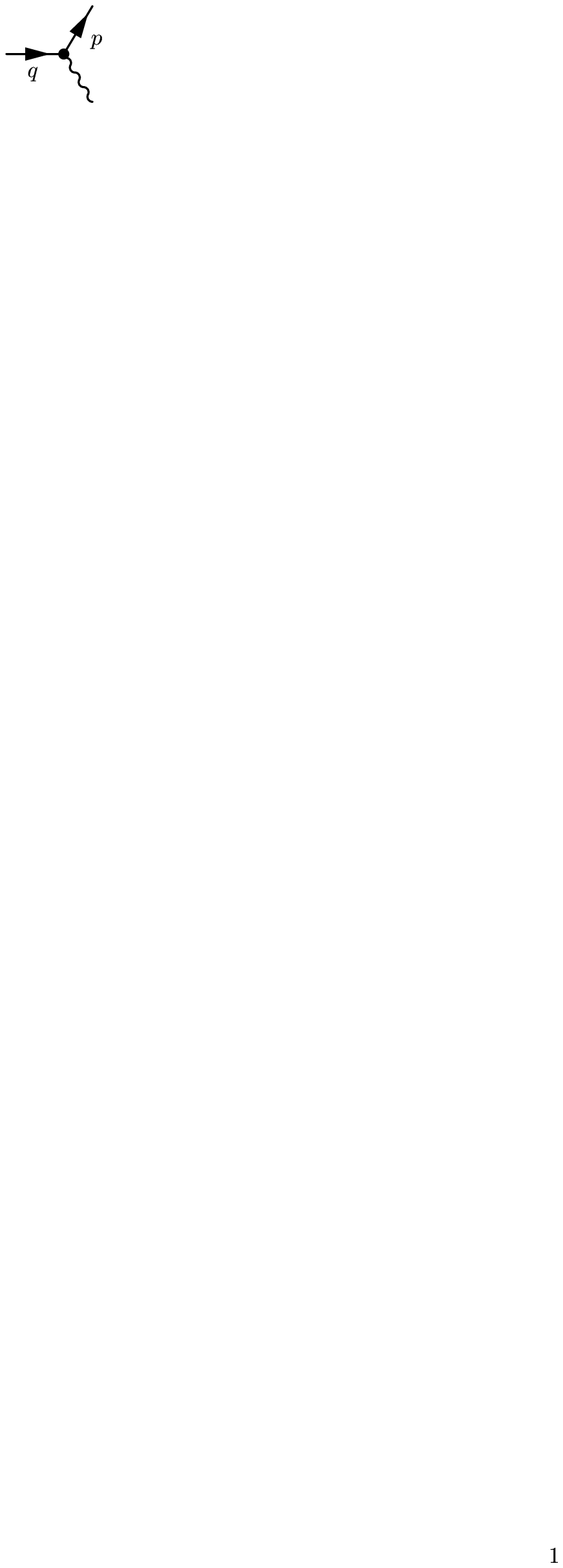}
\end{minipage}
=\Gamma^{\nu} (q,p) = (q^{\nu} + p^{\nu}) F_{K^+} ((p-q)^2) \; .
\end{equation}
For the purpose of the present investigation it is sufficient to parameterize
the kaon charge form factor with a monopole such that the phenomenological
value of the kaon radius, $\langle r_{K^+}^2 \rangle = (0.34 \pm 0.05)$fm$^2$,
is reproduced:
$F_{K^+} (Q^2) = 1/(1+Q^2/(0.84{\rm GeV})^2)$.

In section~\ref{relevance} we have already reflected on the Bethe--Salpeter 
formalism in Euclidean space. We choose the loop momentum $p$ 
to be real, thus the temporal components of the external momenta 
in eqs~(\ref{kapho.extmom.1})--(\ref{kapho.extmom.4}) are purely 
imaginary. Hence the relative momentum $p_f$ as given 
in (\ref{A1.mom.1}) is complex. Since the solution of the Bethe--Salpeter 
equation provides the vertex function $\Phi_{\Lambda}$ only for real 
relative momenta $p_f$, we have to extrapolate $\Phi_{\Lambda}$ to
complex momenta. We fit rational functions to the vertex functions that
are known at $N$ real meshpoints. These rational functions can then
easily be analytically continued. For real momenta a comparison of the
fitted parameterization to the known results allows us to estimate the 
reliability of this treatment.

The differential cross section depends only on the energy $E$ of the 
incoming photon and the angle $\theta$ between the spatial photon 
and kaon momenta. That is illustrated in figure~\ref{kapho_kinme}.
The differential cross section is defined with 
respect to the solid angle element $d\Omega_K=2\pi d(\cos\theta)$ 
of the outgoing kaon:
\begin{equation}\label{kapho.diffcross}
  \frac{d\sigma}{d\Omega_K} (E,\theta) =
       \frac{1}{4} \sum_{s_i,s_f}
       \tilde{\frac{d\sigma}{d\Omega_K}} (E,\theta)
\end{equation}
with
\begin{equation}
  \tilde{\frac{d\sigma}{d\Omega_K}} (E,\theta) =
  \frac{\alpha}{64\pi^2} 4 M_P M_{\Lambda}
  \left| A_1 + A_2 + A_3 \right|^2\, .
\end{equation}
We average, respectively sum over the spins $s_i,s_f$ of the initial
and final states. The phase space factors denoted by $\alpha$ are 
given as
\begin{equation}\label{phasefak.alpha}
  \alpha = \frac{1}{P \cdot p_{\gamma}}
  \frac{| \vec{p}_K |^2}{E_K E_{\Lambda}}
  \left|\frac{d\, |\vec{p}_K|}{d(E_K+E_{\Lambda})}\right|
\end{equation}
with
\begin{equation}\label{phasefak.det}
\left|\frac{d\, |\vec{p}_K|}{d(E_K+E_{\Lambda})} \right| =
\left(\frac{|\vec{p}_K|}{E_K}-\
\frac{\hat{p}_K \cdot \vec{P}_{\Lambda}}
{E_{\Lambda}}\right)^{-1}\, .
\end{equation}
Note that the right hand side of eq~(\ref{phasefak.det}) remains positive
given that $m_{\Lambda} >M_K$. In the CMS the expression
(\ref{phasefak.alpha}) for $\alpha$ simplifies to
\begin{equation}
  \alpha = \frac{| \vec{p}_K |}{| \vec{p}_{\gamma} |}
           \frac{1}{s}\, ,
\end{equation}
where $s=(P + p_{\gamma})^2 = (p_K + P_{\Lambda})^2$ denotes the total
center of mass energy squared. In obtaining the phase space 
factors~(\ref{phasefak.alpha}) we adopted the one particle 
normalization conditions (for Minkowski space)
\begin{equation}
  \langle p | p' \rangle_B = 2p^0 (2\pi)^3 \delta^3 (\vec{p} - \vec{p'})
\quad {\rm and}\quad
  \langle p | p' \rangle_F = \frac{p^0}{m}
        (2\pi)^3 \delta^3 (\vec{p} - \vec{p'}) \, ,
\label{onePartNormF}
\end{equation}
for bosons ($B$) and fermions ($F$). These conventions also enter
the calculation of the transition amplitudes $A_i, i=1,2,3$ and the 
normalization of the Bethe--Salpeter wave--functions.

We obtain the various asymmetries by restricting the sum in 
eq~(\ref{kapho.diffcross}) over the spins to two of the three 
non--scalar particles. We thus obtain the 
$\Lambda$--\emph{polarization asymmetry}
\begin{equation}
\label{p_asym}
  P (E,\theta) = \frac{1}{4}\sum_{s_{p},s_{\gamma}}
     \frac{[s_{\Lambda}=\uparrow] - [s_{\Lambda}=\downarrow]}
     {[s_{\Lambda}=\uparrow] + [s_{\Lambda}=\downarrow]}
\end{equation}
the \emph{polarized target asymmetry}
\begin{equation}
\label{t_asym}
  T (E,\theta) = \frac{1}{2}\sum_{s_{\Lambda},s_{\gamma}}
     \frac{[s_{p}=\uparrow] - [s_{p}=\downarrow]}
     {[s_{p}=\uparrow] + [s_{p}=\downarrow]}           
\end{equation}
and the \emph{polarized photon asymmetry}
\begin{equation}
\label{s_asym}
  \Sigma (E,\theta) =  \frac{1}{2}\sum_{s_{\Lambda},s_{p}}
    \frac{[s_{\gamma}=\uparrow] - [s_{\gamma}=\downarrow]}
    {[s_{\gamma}=\uparrow] + [s_{\gamma}=\downarrow]} \, ,
\end{equation}
where we used the shorthand notation
\begin{equation}
  [s_{\Lambda}=\uparrow] = \left. \tilde{\frac{d\sigma}{d\Omega_K}}
                           \right|_{s_{\Lambda}=\uparrow}
\qquad {\rm etc.}\, .
\end{equation}
Furthermore we denote the spins of the photon, the proton and the 
$\Lambda$ by $s_{\gamma},s_p$ and $s_{\Lambda}$, respectively.

The total cross section is finally obtained from 
eq~(\ref{kapho.diffcross}) via 
\begin{equation}
  \sigma (E) = \int_0^{\pi} \!\! \sin{\theta} \,\, 
               d\theta\frac{d\sigma}{d\Omega_K} (E,\theta)\, .
\end{equation}

\subsection{Associated Strangeness Production  $pp \rightarrow pK\Lambda$}
\label{ap_ASPRO}

\begin{figure}[t] 
\begin{center}
  \epsfig{file=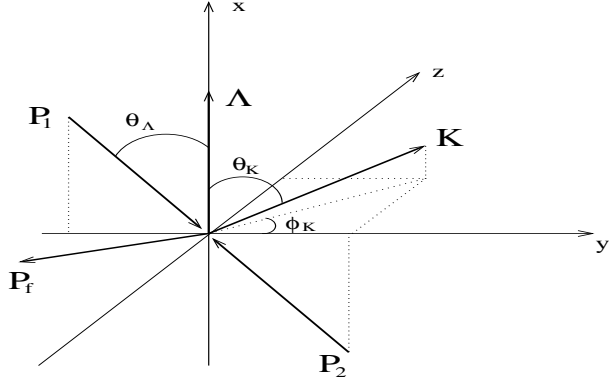,width=8cm,height=5cm}
\end{center}
\caption{\sf Kinematical variables for
    the reaction $pp \rightarrow pK\Lambda$ in the
    center of momentum system.}
\label{pppkl-kin.text} 
\end{figure}

Again we calculate the cross section for the productions process
$pp \rightarrow pK\Lambda$ in the center of momentum frame. The
kinematical setup is depicted in figure~(\ref{pppkl-kin.text}) and
amounts to the momentum routing
%
%
%
%
\begin{eqnarray}
P_1&=&\bigl(\nrmv{p}\cos(\theta_\Lambda), 
\nrmv{p}\sin(\theta_\Lambda),0,\imag E\bigr) \, ,\quad
P_2=\bigl( -\nrmv{p}\cos(\theta_\Lambda),-\nrmv{p}
\sin(\theta_\Lambda),0,\imag E \bigr)\, ,
\nonumber \\
K&=&\bigl( K_x, \nrmv{K} \, \sin(\theta_K) \, \cos(\phi_K), 
\nrmv{K} \sin(\theta_K)\sin(\phi_K),\imag E_K\bigr)\, ,
\nonumber \\
\Lambda &=&\bigl( \Lambda_x,0,0, \imag E_{\Lambda} \bigr) 
 \, ,\quad
P_f =\bigl( \vec{P}_f, \imag E_P \bigr)\, ,
\end{eqnarray}
where $E$ denotes the center of momentum energy. We have furthermore
introduced
\begin{eqnarray}
E_P       &=& 2 E - E_\Lambda - E_K
\, , \quad
\Lambda_x = \sqrt{E_{\Lambda}^2-M_{\Lambda}^2}
\, , \quad
\nrmv{K}  = \sqrt{E_K^2-M_K^2} 
\, , \quad
\vec{P}_f = -\vec{K}-\vec{\Lambda} 
\nonumber\\
K_x &=& \frac{E_P^2-E_K^2-\Lambda_x^2+M_K^2-M_P^2}{2\Lambda_x}
\, , \quad
\sin(\theta_K)=\sqrt{1-\left(\frac{K_x}{\nrmv{K}}\right)^2}\, ,
\end{eqnarray}
where all the momenta are Euclidean.
There are now four independent variables left:
$E_\Lambda$, $E_K$, $\theta_\Lambda$ and $\phi_K$.
As for $p\gamma\to K\Lambda$ we find the vertex function 
$\Gamma_\Lambda$ for complex arguments by analyticly continuing
a fitted rational function.

The differential cross section is given by
\begin{equation}
\frac{d\sigma}{d\Omega_\Lambda} 
=\frac{1}{4} \sum_{s,s',r,r'}
\int\, d\Sigma\, |{\mathcal M}_{s,s',r,r'}|^2 \, ,
\quad {\rm where}\quad
d\Sigma= \frac{1}{128 \pi^5}\frac{M_P^3 M_\Lambda}{|\vec{p}_{cm}|E}
dE_K\, dE_{\Lambda}\, d\phi_K \, ,
\label{ap_as_diffcross}
\end{equation}
with $d\Omega_\Lambda = 2\pi \, d{\rm cos}(\theta_\Lambda)$. The masses
of the proton and the $\Lambda$ are denoted by $M_P$ and $M_{\Lambda}$
while $(\vec{p}_{cm},\imag E)$ represents the four vector of the total 
momentum in the CMS. The amplitude ${\mathcal M}_{s,s',r,r'}$ for the 
reaction $p\gamma \to K\Lambda$ depends on the spin orientation of
the incoming and outgoing particles. In (\ref{ap_as_diffcross}) we
average, respectively sum over the spin projections of the incoming and 
outgoing particles. The integrations over the kaon and $\Lambda$ energies 
are constrained by the available energy, which is $2E$.

For each diagram in figure~\ref{pppkldiag} and for each specific diquark 
content ({\it cf}. Table \ref{tab_AsStr_1}) one has a contribution of the
form
\begin{equation}
{\mathcal L} \, \left( \frac{i}{Q^2+m_{\phi}^2} \right) \, {\mathcal H}
\end{equation}
to the amplitude $\mathcal{M}$. Here $m_\phi$ refers to the mass
of the intermediate pseudoscalar meson, {\it i.e.} $m_\phi=M_\pi$
or $m_\phi=M_K$. The `meson matrix element' is given by 
\begin{equation}  
{\cal L}_{s,s'} = \bar{u}_{s'}(P_2) \: i\gamma_5 \, 
g_{\pi NN}(Q^2)\: u_s(P_f) 
\quad {\rm or} \quad
{\cal L}_{s,s'} = \bar{u}_{s'}(P_2) \: 
i\gamma_5 \, g_{KN\Lambda}(Q^2)\: u_s(P_\Lambda)\, ,
\end{equation}
depending on whether the intermediate meson is a pion or a kaon.
The `handbag part' ${\mathcal H}$ has the general structure
\begin{equation}
{\mathcal H}=i\int\frac{d^4p}{(2\pi)^4}\left\{\bar{\Phi}_{\Lambda}(p_f,P_f)\,
S(p_+)\,\Gamma_K(p_+,q)\,S(q)\right\} 
\left\{\Gamma_\pi(q, p_-)\,S(p_-)D(p_d)\,\Phi_{P}(p_i,P_i)\right\}.
\end{equation}
Here the Bethe--Salpeter amplitudes $\bar{\Phi}_{\Lambda}$ and $\Phi_{P}$ 
for the $\Lambda$ and the proton as well as the (di)quark propagators
$S$ and $D$ enter. The meson--quark vertices $\Gamma_{\pi}$ and $\Gamma_K$ 
are defined in eq~(\ref{qKVertex}). 

The calculation of the amplitudes involves four dimensional loop
integrations. All the momenta are Euclidean and therefore we
use hyperspherical coordinates. For the inner loop of the handbag 
diagrams we use a Gauss--Legendre routine, whereas the phase space 
integration is performed with Monte--Carlo methods. Due to the 
considerable effort it takes to integrate eight integrals numerically 
the calculation could only be performed to an overall accuracy of 
5\% to 15\%. However, we consider that sufficient for a comparison 
with data.  

The depolarization tensor is defined as
\begin{equation}
 D_{NN}(x_F) = \frac{a-b}{a+b}  
\end{equation}
with the shorthand notation
\begin{equation}
a=\frac{d\sigma}{dx_F}
(\uparrow_p\uparrow_{\Lambda}+\downarrow_p\downarrow_{\Lambda}) 
\quad {\rm and} \quad
b=\frac{d\sigma}{dx_F}
(\uparrow_p\downarrow_{\Lambda}+\downarrow_p\uparrow_{\Lambda})
\end{equation}
for the cross sections with different spin projections.
Here $x_F$ denotes the real momentum of the $\Lambda$ scaled by the
maximum value allowed by the kinematics:
\begin{equation}
x_F = \frac{|\vec{\mathbf{\Lambda}}_\||}
{|\vec{\mathbf{\Lambda}}_{\|max}|}\, .
\end{equation}

\end{appendix}

\end{document}